\documentclass[preprint]{aastex}

\newcommand\spirtfr{\hbox{1}}
\newcommand\lumcuts{\hbox{2}}
\newcommand\fulltfr{\hbox{3}}

\shorttitle{Scatter in the Tully-Fisher Relation}
\shortauthors{Kannappan, Fabricant, \& Franx}
 
\begin{document}

\title{Physical Sources of Scatter in the Tully-Fisher Relation}

\author{Sheila J. Kannappan\altaffilmark{1,2},
Daniel G. Fabricant\altaffilmark{1}, and Marijn Franx\altaffilmark{3}}

\altaffiltext{1}{Harvard-Smithsonian Center for
Astrophysics, 60 Garden St. MS-20, Cambridge, MA 02138;
skannappan@cfa.harvard.edu, dfabricant@cfa.harvard.edu}

\altaffiltext{2}{Current
address: The University of Texas at Austin, McDonald Observatory, 2511
Speedway, RLM 15.308 C1400, Austin, TX 78712; sheila@astro.as.utexas.edu}

\altaffiltext{3}{Leiden Observatory, PO
Box 9513, 2300 RA Leiden, Netherlands; franx@strw.leidenuniv.nl}

\begin{abstract}
We analyze residuals from the Tully-Fisher relation for the
emission-line galaxies in the Nearby Field Galaxy Survey, a broadly
representative survey designed to fairly sample the variety of galaxy
morphologies and environments in the local universe for luminosities
from $\rm M_{\rm B}=-15$ to $-23$.  For a subsample consisting of the
spiral galaxies brighter than $\rm M_{\rm R}^i=-18$, we find strong
correlations between Tully-Fisher residuals and both $B-R$ color and
EW(H$\alpha$).  The extremes of the correlations are populated by Sa
galaxies, which show consistently red colors, and spiral galaxies with
morphological peculiarities, which are often blue.  If we apply an
EW(H$\alpha$)-dependent or $B-R$ color-dependent correction term to
the Tully-Fisher relation, the scatter in the relation no longer
increases from R to B to U but instead drops to a nearly constant
level in all bands, close to the scatter we expect from measurement
errors.  We argue that these results probably reflect correlated
offsets in luminosity and color as a function of star formation
history.  Broadening the sample in morphology and luminosity, we find
that most non-spiral galaxies brighter than M$_{\rm R}^i=-18$ follow
the same correlations between Tully-Fisher residuals and $B-R$ color
and EW(H$\alpha$) as do spirals, albeit with greater scatter.
However, the color and EW(H$\alpha$) correlations do not apply to
galaxies fainter than M$_{\rm R}^i=-18$ or to emission-line S0
galaxies with anomalous gas kinematics.  For the dwarf galaxy
population, the parameters controlling Tully-Fisher residuals are
instead related to the degree of recent evolutionary disturbance:
overluminous dwarfs have higher rotation curve asymmetries, brighter
U-band effective surface brightnesses, and shorter gas consumption
timescales than their underluminous counterparts.  As a result, sample
selection strongly affects the measured faint-end slope of the
Tully-Fisher relation, and a sample limited to include only passively
evolving, rotationally supported galaxies displays a break toward
steeper slope at low luminosities.
\end{abstract}


\section{Introduction}
\label{sc:introtf1}

The tight correlation between luminosity and rotation velocity for spiral
galaxies, a.k.a.\ the Tully-Fisher relation
\citep[TFR,][]{tully.fisher:new}, has motivated dozens of studies over the
years.  In pursuit of the best possible distance indicator, most TF
analyses have been carried out in red or infrared passbands \citep[see
reviews by][]{strauss.willick:density,jacoby.branch.ea:critical}, which
offer lower scatter than bluer passbands.  However, recent theoretical work
has emphasized the value of studying TF scatter for its own sake, as this
scatter holds fundamental clues to the formation and evolution of galaxies
\citep[e.g.][]{buchalter.jimenez.ea:galactosynthesis,elizondo.yepes.ea:self-regulating,eisenstein.loeb:can}.
Observations in bluer passbands are extremely useful for understanding TF
scatter, because the scatter arising from recent star formation,
differences in stellar populations, and the spread in formation redshifts
is most visible in the blue.

Physical scatter in the TFR can come from three sources: (1)
variations in the stellar mass-to-light ratio, (2) variations in the
stellar mass fraction (stellar-to-total mass ratio), and (3)
differences in how the observed velocity width relates to the total
mass.  By identifying physical scatter from any one of these sources
we limit the contributions of the other two, simultaneously gaining
insight into the star formation histories of galaxies, the
relationship between visible galaxies and dark matter halos, and the
dynamics and structure of galaxy disks.

A better understanding of the sources of TF scatter may also minimize
uncertainties in other types of TF analyses.  For example, if TF slope
varies with environment, or if galaxies with different colors define
different TF zero points, then a cluster TFR calibrated with a small set of
galaxies that have Cepheid distances may not be universal \citep[e.g.\ as
in][]{tully.pierce:distances,sakai.mould.ea:hubble}. Correcting for these
systematics could reduce scatter and improve the TFR as a distance
indicator.  Calibrating zero point shifts as a function of color and
emission line strength could also aid in the interpretation of evidence for
luminosity evolution in the intermediate-redshift TFR \citep[][see also
Kannappan, Fabricant, \& Franx, in
preparation]{simard.pritchet:internal,vogt.phillips.ea:optical}.

Here, we use a well-defined sample of emission-line galaxies drawn
from the 196-galaxy Nearby Field Galaxy Survey \citep[NFGS,
][]{jansen.franx.ea:surface,kannappan:kinematic} to
examine how offsets from the U, B, and R-band TFRs depend upon the
physical properties of galaxies.  While previous TF studies have often
included hundreds or thousands of galaxies
\citep[e.g.][]{aaronson.huchra.ea:velocity,pierce.tully:luminosity-line,mathewson.ford.ea:southern,willick.courteau.ea:homogeneous,giovanelli.haynes.ea:i*1,courteau:optical,sakai.mould.ea:hubble,tully.pierce:distances},
none of these data sets has both the broadly representative sample
demographics and the wide array of supporting data of the NFGS
(\S\ref{sc:nfgstf1}).  Thus despite its modest size, the present survey
offers unique advantages for studying scatter in the TFR.

The bulk of this paper analyzes a sample of NFGS galaxies consisting
of $\sim$70 Sa--Sd spiral galaxies brighter than M$_{\rm R}^i=-18$.  This
spiral sample allows us to investigate the behavior of galaxies often
excluded from the TFR: Sa galaxies and galaxies with peculiarities
such as warps, multiple nuclei, or interacting companions.
Ultimately, we seek to fit these galaxies into a unified picture, in
which we understand TF residuals in terms of continuous physical
properties.  To this end we undertake a detailed analysis of
third-parameter correlations with TF residuals, and we attempt to form
a physical understanding of TF scatter.

The remainder of the paper draws from the full NFGS to define an
extended sample of $\sim$110 galaxies including very late types,
emission-line E/S0 galaxies, and faint dwarfs.  The extended sample
allows us to search for physical drivers of TF scatter in a
heterogeneous population.  The only other comparable TF sample is the
$\sim$40-galaxy Ursa Major sample of \citet{verheijen.sancisi:ursa},
which was drawn from a restricted environment.  The Virgo and Fornax
cluster samples of \citet{yasuda.fukugita.ea:study} and
\citet{bureau.mould.ea:new} apply morphology restrictions that exclude
irregular or interacting galaxies.  Although some intermediate-redshift TF
analyses have included a wide range of morphologies
\citep{rix.guhathakurta.ea:internal,forbes.phillips.ea:keck,simard.pritchet:internal},
these studies nonetheless favor a select population: typically,
moderately bright blue galaxies with strong emission and high surface
brightness.  And although the samples studied by
\citet{pierini.tuffs:linear} and \citet{mcgaugh.schombert.ea:baryonic}
both span large ranges in luminosity, the former study is restricted
by Hubble type and evidence for interaction, while the latter study is
drawn from a heterogeneous mix of smaller samples selected by various
criteria in different photometric bands.  The NFGS thus provides an
unusually broad and unbiased representation of the general galaxy
population.

In what follows, we describe the survey and briefly review our data
reduction and analysis techniques (\S\ref{sc:nfgstf1}--\ref{sc:ldata}),
then present results for the spiral sample
(\S\ref{sc:spiralcal}--\ref{sc:spiral3}) and for the extended sample
(\S\ref{sc:fulltfr}).  We summarize our major conclusions in
\S\ref{sc:concltf1}.  Further details on our data analysis methods may be
found in Appendices~\ref{sc:fitting}--\ref{sc:asymmeast}.

\section{The Nearby Field Galaxy Survey}
\label{sc:nfgstf1}

The comprehensive nature of the Nearby Field Galaxy Survey (NFGS) enables
us to analyze how TF scatter depends on a broad range of galaxy properties.
The database includes UBR photometry, integrated and nuclear
spectrophotometry, and gas and stellar kinematic data for a statistically
representative sample of the local galaxy population \citep[][Kannappan et
al., in preparation]{jansen.fabricant.ea:spectrophotometry,
jansen.franx.ea:surface}.  The 196 NFGS galaxies were objectively selected
from the CfA~1 redshift survey \citep{huchra.davis.ea:survey} without
preference for morphology, color, diameter, inclination, environment, or
any other galaxy property, as described in \citet{jansen.franx.ea:surface}.
To counteract the bright galaxy bias of CfA~1, galaxies were included
approximately in proportion to the local B-band galaxy luminosity function
\citep[LF, e.g.][]{marzke.huchra.ea:luminosity}, with the final sample
spanning luminosities from M$_{\rm B}$ $\sim$$-$23 to $-$15.

\begin{figure}[tb]
\epsscale{.5}
\plotone{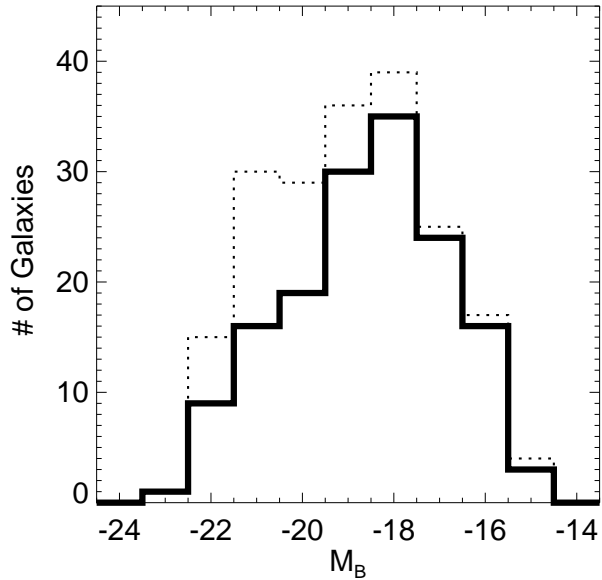}
\caption{Luminosity distribution for the Nearby Field Galaxy Survey,
using B-band CCD magnitudes from the NFGS database
\citep{jansen.franx.ea:surface} converted to H$_{0}$=75.  The dotted
line shows the full NFGS and the solid line shows the 153 galaxies
with extended ionized gas emission (optical rotation curves).}
\label{fg:lf}
\end{figure}

The resulting sample provides an unusually unbiased sampling of the general
galaxy population.  However, no sample is completely free of selection
effects.  The NFGS is subject to the inherent color and surface brightness
biases of the parent CfA~1 survey, itself a descendent of the Zwicky
catalog and therefore essentially B-selected.  Also, despite the attempt to
favor faint galaxies in the selection process, the NFGS reproduces the
local LF imperfectly.  Figure~\ref{fg:lf} shows that the sample
luminosity distribution varies slowly over the range $-16>\rm M_{\rm
B}>-22$ and cuts off for brighter and fainter galaxies.

Another potential concern is the luminosity-distance correlation built into
the NFGS.  The survey was selected with a luminosity-dependent lower
redshift limit instead of a maximum diameter limit, in order to increase
the odds that galaxies would fit within the length of the 3$\arcmin$
spectrograph slit without biasing the sample in diameter
\citep{jansen.franx.ea:surface}.  As a result, fainter galaxies have
greater fractional uncertainties in distances and absolute magnitudes, and
such galaxies receive less weight in error-weighted TF fits.  For this
reason, we adopt unweighted fits as our primary fitting technique (see
Appendix~\ref{sc:fitting} for a discussion of the differences between
techniques).

\section{Velocity Width Data}
\label{sc:vdata}

\subsection{Velocity Width Sample}

\begin{figure}[tb]
\epsscale{0.6}
\plotone{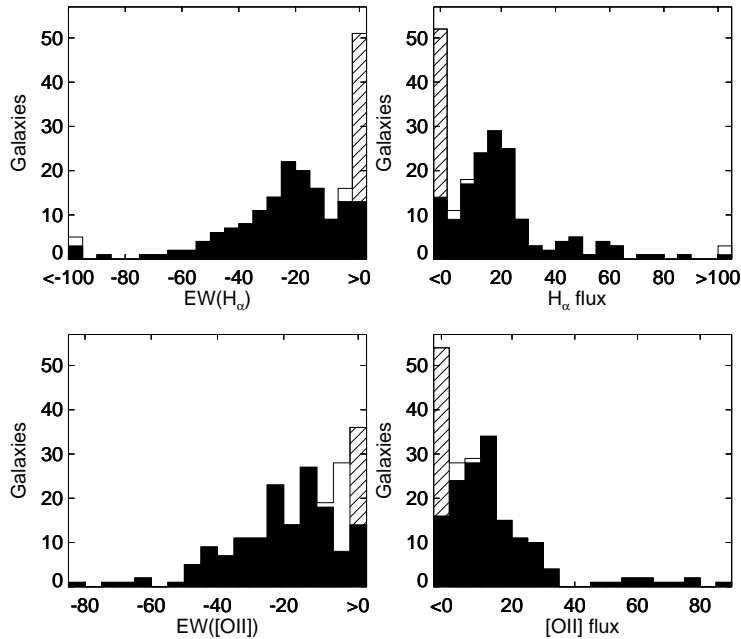}
\caption{Detection of extended emission (rotation curves) as a function of
emission line strength.  Left panels: H$\alpha$ and [OII] 3727 equivalent
widths for the entire NFGS (outline) and for the 153-galaxy optical RC
sample (solid).  Right panels: Calibrated line fluxes in the same format.
Cross-hatched bins represent all galaxies for which the integrated
spectrophotometry indicates lines in absorption; with the higher resolution
kinematic data we can detect extended emission in a number of these
galaxies.  The scales of the top two panels are truncated for clarity.
Galaxies with EW(H$\alpha$) less than $-$100 are plotted in the first bin
of the top left panel; those with H$\alpha$ flux greater than 100 are
plotted in the last bin of the top right panel.  Galaxies with strong
H$\alpha$ emission but no rotation curve are AGNs.  We use equivalent
widths from the NFGS database \citep{jansen.fabricant.ea:spectrophotometry}
and fluxes courtesy of R. A. Jansen (private communication).}
\label{fg:emhists}
\end{figure}

Our primary velocity widths are derived from the full set of 153
major-axis ionized gas rotation curves (RCs) obtained for the NFGS
kinematic database (Kannappan et al., in preparation).  This
sample is complete in the sense that kinematic observations were
attempted for all NFGS galaxies; however, emission line detection
limits varied with observing conditions, integration times, and
available emission lines (see \S\ref{sc:orcobs}).
Figure~\ref{fg:emhists} displays the emission line properties of the
optical RC sample in the context of the NFGS as a whole.

\begin{figure}[tb]
\epsscale{0.6}
\plotone{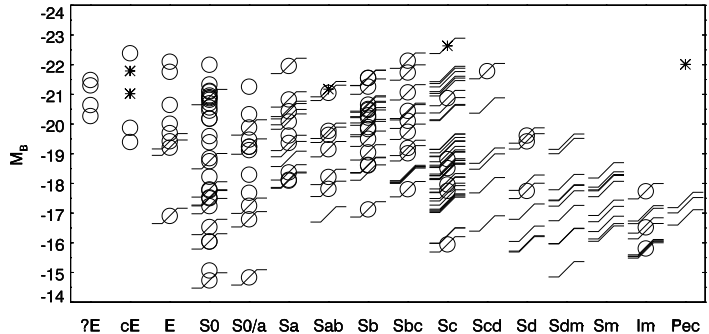}
\caption{Overview of the Nearby Field Galaxy Survey, showing the
demographics of all 196 galaxies sorted by morphology and B-band
luminosity.  A circular symbol indicates that we have stellar
absorption line data for the galaxy, while an S-shaped symbol
indicates that we have extended gas emission line data.  The optical
RC sample consists of the 153 galaxies shown with an S-shaped symbol.
Strong AGNs are marked with a star.}
\label{fg:typemagdistrib}
\end{figure}

As illustrated in Figure~\ref{fg:typemagdistrib}, the optical RC sample
includes all NFGS galaxies of type Sab and later (including type ``Pec''
galaxies), except for two galaxies that show largely featureless spectra.
In addition, the sample includes 12 of the 13 Sa's in the NFGS and 17 of
the E/S0's.

For comparison with the optical data, we have also extracted
H\,{\small I} 21-cm linewidths for 105 NFGS galaxies from the catalogs
of \citet{bottinelli.gouguenheim.ea:extragalactic} and
\citet{theureau.bottinelli.ea:kinematics}.  These galaxies are not a
statistically fair subsample of the NFGS: H\,{\small I} data are more
often missing for the higher luminosity NFGS galaxies.  We exclude from
analysis one galaxy that has H\,{\small I} data but no optical RC
(NGC~2692, an Sa galaxy).

The NFGS includes galaxies at all inclinations $i$, but $\sin{i}$
corrections are uncertain for galaxies with $i<40$, so most of our
analysis considers only the 108 NFGS emission-line galaxies with
$i>40$.  However, we do not restrict inclinations when calibrating
optical-to-radio conversions (\S\ref{sc:vcalcs}), because no
inclination correction is required.

\subsection{Optical Rotation Curve Observations}
\label{sc:orcobs}

Long slit spectra were obtained during several observing runs between 1996
and 1999, using the FAST spectrograph on the Tillinghast Telescope
\citep{fabricant.cheimets.ea:fast}.  For most galaxies we observed the
H$\alpha$, [NII], and [SII] lines between $\sim$6200--7200 \AA, with
spectral resolution $\sigma\sim $ 30 $\rm km \, s^{-1}$, a
2$^{\prime\prime}$ or occasionally 3$^{\prime\prime}$ wide slit, and a
spatial binning of 2.3$^{\prime\prime}$/pixel (comparable to the typical
seeing of 2$^{\prime\prime}$).  For some galaxies, in conjunction with
stellar absorption line observations, we also observed the H$\beta$ and
[OIII] emission lines between $\sim$4100--6100 \AA, again with
2.3$^{\prime\prime}$ spatial binning and a 2$^{\prime\prime}$ slit, but
with reduced spectral resolution, $\sigma\sim 60$ $\rm km \, s^{-1}$.
These lower resolution observations serve as the primary emission line data
for 10 galaxies for which we lack high resolution observations.

The alignment between the slit position angle and the galaxy major
axis position angle listed in the UGC \citep{nilson:uppsala} was
generally tight, $\Delta$P.A. $<$ 6$\degr$.  For two galaxies the slit
was severely misaligned ($\Delta$P.A. = 40$\degr$ and 75$\degr$).
Also, a few galaxies had no UGC P.A. and were instead assigned P.A.'s
based on visual inspection of the digitized POSS images; these
galaxies have uncertain $\Delta$P.A.  All galaxies with large or
uncertain $\Delta$P.A. have photometric inclination $i<40$ and so are
automatically excluded from our TF samples.  In Appendix~\ref{sc:vdefns},
where we do not employ an inclination restriction, we explicitly
exclude galaxies with large or uncertain $\Delta$P.A.

All of the data were reduced using standard methods, including bias and
dark subtraction, flat-fielding, wavelength calibration, heliocentric
velocity correction, sky subtraction, spectral straightening, and cosmic
ray removal, using IRAF and IDL.  We extracted rotation curves by fitting
all available emission lines simultaneously with fixed line spacing in
$\log{\lambda}$, rejecting fits with S/N $\la$ 3.  In cases of severe
H$\alpha$ or H$\beta$ absorption, these lines were omitted from the fits.

\subsection{Velocity Width Definitions}
\label{sc:vcalcs}

We have considered three possible optical velocity measures: $V_{max}$, the
single largest velocity in the rotation curve; $V_{fit}$, the velocity
interpolated at the ``critical radius'' \citep[1.3$r_e$, the peak velocity
position for a theoretical exponential disk,][]{freeman:on} using a
functional fit to the rotation curve \citep[cf.][]{courteau:optical}; and
$V_{pmm}$, defined as half the difference between the statistical
``probable minimum'' and ``probable maximum'' velocities in the rotation
curve \citep[cf.][]{raychaudhury..ea:tests}.  The prescriptions for these
velocity measures are fully described in Appendix~\ref{sc:vdefns}.  As
illustrated in Figure~\ref{fg:rcgallery}, $V_{max}$ is overly sensitive to
details of RC shape.  $V_{fit}$ sucessfully models bright spiral galaxies
similar to those in Courteau's sample but sometimes fails when confronted
with the full variety of rotation curve shapes in the NFGS.  For our
sample, the most robust velocity width estimator is $V_{pmm}$, which makes
use of all of the data without imposing a specific model on the RC.

\begin{figure}[tb]
\epsscale{.5}
\plotone{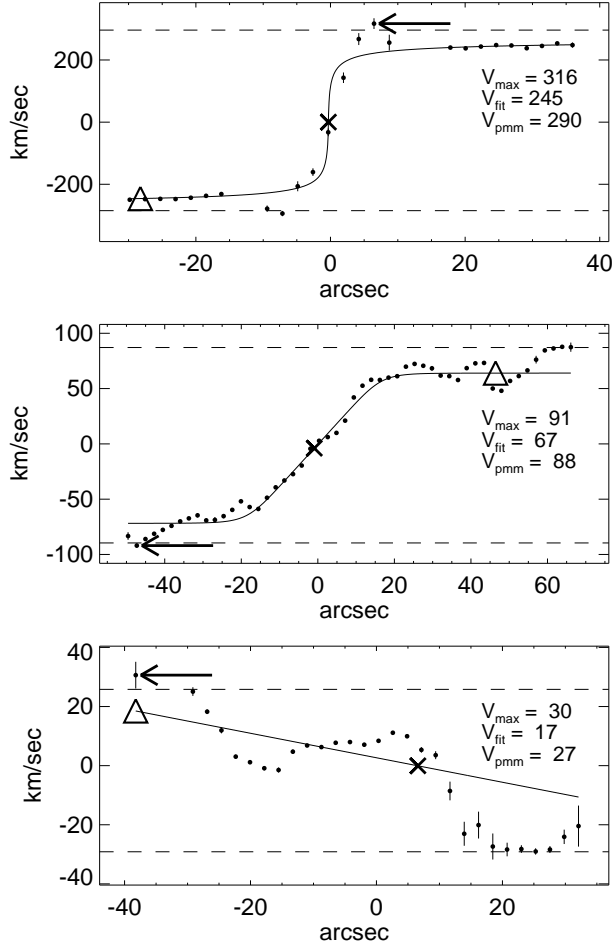}
\caption{Comparison of optical velocity width definitions (described fully
in Appendix~\ref{sc:vdefns}).  The data are shifted to the galaxy rest
frame using the redshift determined by minimizing RC asymmetry
(Appendix~\ref{sc:asymmeast}), and the arrow marks the largest velocity
with respect to that origin, defined as $V_{max}$.  The two dashed lines
indicate the probable minimum and maximum velocities, with half the
difference between the two equalling $V_{pmm}$ (independent of origin).
The solid line shows a functional fit to the data with the RC origin varied
as a fit parameter; an X marks the fitted origin and an open triangle marks
the point defined as $V_{fit}$, which is interpolated at 1.3$r_e$ where
$r_e$ is the B-band half-light radius along the major axis \citep[converted
from the elliptical $r_e$ in][]{jansen.franx.ea:surface}.}
\label{fg:rcgallery}
\end{figure}

Except when comparing our data with other optical RC TF samples, we convert
all of our optical and radio velocity widths to an equivalent W$_{50}$
linewidth scale \citep[where W$_{p}$ is the HI linewidth at $p$\% of peak
intensity, see][]{bottinelli.gouguenheim.ea:extragalactic}.  For galaxies
with W$_{20}$ but no W$_{50}$ in the catalog, we simply subtract 20 $\rm km
\,s^{-1}$ from the W$_{20}$ linewidths and use the adjusted W$_{20}$'s
together with the W$_{50}$'s, referring to the combined data set as W$_{\rm
HI}$.  The number 20 $\rm km \,s^{-1}$ comes from a comparison of W$_{20}$
and W$_{50}$ where we have both measures;
\citet{haynes.giovanelli.ea:i-band} derive a similar value from a much
larger sample.

Appendix~\ref{sc:vdefns} presents the empirically fitted relations we use
to convert raw optical $V$'s to W$_{50}$-equivalent W$_{V}$'s, as
determined from the 96 galaxies for which we have both optical RC and
W$_{50}$ data.  We use the scatter in the conversion relations to set the
nominal error bars for the final velocity widths.  These error bars include
not only the optical RC measurement errors (which are small) but also all
other sources of discrepancy between optical and radio velocity widths,
including P.A. misalignment errors as well as ``physical'' errors due to
fundamental differences between the kinematics of unresolved global HI
fields and the kinematics of HII regions that happen to lie along arbitrary
1D lines.  {\em However, contrary to common assumption, the limited extent
of the optical RCs does not lead to any significant discrepancy with the
global HI linewidths.}  Radio W$_{50}$'s and optical velocity widths agree
well even for the NFGS galaxies whose RCs are truncated at $<$1.3$r_e$ (see
Appendix~\ref{sc:vdefns}).  Except when comparing with
\citet{courteau:optical}, we adopt the W$_{50}$-equivalent velocity width
W$_{V_{pmm}}$ as our default optical velocity width, with a nominal
error bar of 20 $\rm km \,s^{-1}$.

\clearpage

\subsection{Corrections to Velocity Widths}
\label{sc:vcorrs}

All optical velocity widths are computed from rest-frame rotation
curves, so no additional correction for cosmological expansion is
necessary.  H\,{\small I} linewidths are corrected for cosmological
expansion by dividing by $(1+z)$.

We correct for inclination by dividing by $\sin i$, denoting corrected
velocity widths with a superscript $i$.  Here
\begin{equation}
i = \cos^{-1} \sqrt {\left( \left( b/a \right) ^2 - q_{0}^2 \right)/ 
\left( 1 - q_{0}^2 \right) }
\end{equation}
with $i$ set to 90 whenever $b/a < q_{0}$, the assumed intrinsic
minor-to-major axis ratio.  We adopt $q_{0}=0.20$ except when comparing
with \citet{courteau:optical}, in which case we use his value (0.18).
Tests with a range of $q_{0}$'s indicate that TF scatter is remarkably
insensitive to the details of computing inclinations.  The errors are
propagated assuming errors in $b$ and $a$ equal to their measurement
precision in the UGC (typically 0.05--0.1$\arcmin$).  We have checked the
quality of the UGC estimates against estimates derived from both 2MASS data
and NFGS CCD photometry (R. Jansen, private communication): the newer
measurements do not improve TF scatter.

We do not apply turbulence corrections to either optical or radio velocity
widths, except when calculating extinctions (see \S\ref{sc:extinct}).
Turbulence estimates are uncertain, and furthermore turbulence corrections
may be inappropriate for small galaxies and non-equilibrium systems.  In
fact turbulence may provide an important source of support for such
systems, and we might wish to {\em add} a gas pressure correction to the
optical velocity widths.  The conversion of optical velocity widths to an
equivalent W$_{50}$ scale (\S\ref{sc:vcalcs}) effectively makes such a
correction, averaged over the sample.

Simulations by \citet{giovanelli.haynes.ea:i} indicate that slit
misalignment effects remain negligible for $\Delta$P.A. $<$ 15$\degr$,
and observational tests in the range $\Delta$P.A. $<$ 8$\degr$ confirm
this result \citep{courteau:optical}.  For galaxies with $i>40$, our
P.A. alignments are all within 6$\degr$ of the reference P.A.'s listed
in the UGC (\S\ref{sc:orcobs}).  Comparison between the UGC P.A.'s and
corresponding automated P.A. measurements from 2MASS and POSS
\citep{garnier.paturel.ea:image} suggests typical catalog-to-catalog
deviations of under 15$\degr$.  Therefore the combined misalignment
from observational mismatch and reference uncertainty should nearly
always be within the range tested by \citet{giovanelli.haynes.ea:i},
and we choose to make no correction.  However, an rms
P.A. misalignment error term is {\em de facto} included in our errors,
because we estimate the optical velocity width errors based on the
scatter in the optical-to-radio conversion (\S\ref{sc:vcalcs}).

\section{Luminosity Data}
\label{sc:ldata}

Extrapolated total B and R magnitudes are available in the NFGS
photometry database \citep{jansen.franx.ea:surface} for all galaxies
with optical RCs.  In addition, \citeauthor{jansen.franx.ea:surface}\
provide U-band magnitudes for all but two of the galaxies.  The
database magnitudes include corrections for galactic absorption and
have typical errors of $\sim$0.02--0.03 mag.

\subsection{Absolute Magnitudes}
\label{sc:absmag}

We compute absolute magnitudes using distances derived from a linear
Hubble flow model with H$_{0}=75$, corrected for Virgocentric infall.
For consistency, we use the redshifts tabulated by
\citet{jansen.franx.ea:surface} and adopt their Virgocentric infall
correction \citep[from][]{kraan-korteweg.sandage.ea:effect}.
Comparison with our newly measured redshifts indicates good agreement
between the two sets of measurements ($\sigma\sim$35 $\rm km \,
s^{-1}$) with a few outliers.  We assign a redshift error of the
greater of 35 $\rm km \, s^{-1}$ or the difference between the
tabulated and newly measured redshifts and add this term in quadrature
with an assumed peculiar velocity uncertainty of 200 $\rm km \,
s^{-1}$.

\subsection{Internal Extinction Corrections}
\label{sc:extinct}

Magnitudes corrected for internal extinction are denoted by a
superscript $i$.  Except when comparing with \citet{courteau:optical},
we adopt the velocity width-dependent (and thus implicitly
luminosity-dependent) internal extinction corrections of
\citet{tully.pierce.ea:global}:
\begin{equation}
A_{i}^{\lambda}=\gamma_{\lambda}\log{\left( a/b \right)},
\end{equation}
where $a/b$ is the major to minor axis ratio and $\gamma_{\lambda}$ is
the extinction coefficient.  \citeauthor{tully.pierce.ea:global}
measure $\gamma_{\lambda}$ at B, R, and K'.  For the U band we use
extrapolated coefficients:
\begin{equation}
\gamma_{U} = 1.69 + 3.00 \left( \log W^{i}_{R} - 2.5 \right).
\end{equation}

Here $W_{R}$ is the velocity width on the turbulence-corrected scale
of \citet{tully.fouque:extragalactic}:
\begin{equation}
W_R^2=W_{20}^2 + W_t^2 - 2 W_t W_{20} \left( 1-e^{-W_{20}^2/W_c^2}
\right) - 2W_t^2 e^{-W_{20}^2/W_c^2} + 4W_{dwarf}^2,
\end{equation}
where $W_t=38$, $W_c=120$, and $W_{dwarf}=17$. Tully \& Fouque
introduce the last term to define a dynamical velocity width suitable
for dwarf galaxies; it prevents small linewidths from being
``corrected'' to square roots of negative numbers.  Note that we add
the 17 $\rm km \, s^{-1}$ ``restored turbulence'' term before the
inclination correction, apparently contrary to the formula in Tully \&
Fouque, because otherwise the turbulence correction is multiplied by
$1/\sin{i}$ while the restoration is not, which still results in
square roots of negative numbers.  While we do not generally use
turbulence corrections (see \S\ref{sc:vcorrs}), for consistency with
Tully et al.\ we make an exception when computing $W_R$.
W$_{20}$-equivalent linewidths are calculated from optical velocity
widths as described in \S\ref{sc:vcalcs} and Appendix~\ref{sc:vdefns}.

We apply the above extinction corrections to galaxies of type Sa and later;
extinction corrections for E/S0 galaxies are set to zero.  Except for one E
galaxy, this choice has little impact, since applying the extinction
formula to the other E/S0's would yield corrections of $\la$0.15 mag.  Note
that we {\em do} apply the correction to dwarf galaxies.  Although the
\citet{tully.pierce.ea:global} extinction corrections were derived for a
sample composed primarily of spiral galaxies, \citet{pierini:internal} has
shown that they perform well for dwarf galaxies.  As prescribed by
\citeauthor{tully.pierce.ea:global}, the extinction corrections are set to
zero whenever they go negative, to avoid having the corrected luminosities
come out fainter than the observed luminosities.  For the same reason, the
error bars on the corrected magnitudes do not extend fainter than the
measurement errors on the observed magnitudes.

\citet{courteau:deep}, following
\citet{willick.courteau.ea:homogeneous}, uses an alternative
correction formula with no velocity width or luminosity
dependence. However, if we follow their procedure and choose a
luminosity-independent value of $\gamma$ to best minimize TF scatter,
we find that no value of $\gamma$ reduces scatter as well as the
\citeauthor{tully.pierce.ea:global}\ formulation.  In the R band, the
\citeauthor{tully.pierce.ea:global}\ corrections perform only
marginally better than the luminosity-independent corrections, but at
B and U the differences become more substantial (at U, scatter
decreases by 18\% compared to 8\%).  We therefore conclude that the
general behavior, if not the zero point, of the luminosity-dependent
corrections appears correct.

\section{The Spiral TFR 1:  Basic Calibration \& Literature Comparison}
\label{sc:spiralcal}

Traditional TF samples consist of moderately bright spiral galaxies,
chosen to be relatively edge-on to avoid uncertain $\sin{i}$
corrections.  In this section and
\S\ref{sc:pecandsa}--\ref{sc:spiral3} we analyze the TFR for spiral
galaxies brighter than $\rm M_{\rm R}^i=-18$ and inclined by more than
40$\degr$.  The NFGS contains 69 Sa--Sd galaxies that meet these
criteria; optical velocity widths and magnitudes are available for 68
of these (67 at U).  H\,{\small I} linewidths are available for 46 of
the 68 (or 45 of the 67).

\subsection{Basic Calibration}
\label{sc:basiccal}

Table~\spirtfr\ gives TF fit parameters for the spiral TF sample in U,
B, and R, including the observed scatter values and the scatter values
predicted to arise from measurement errors.  We provide fit results
using three different types of fits, as discussed in
Appendix~\ref{sc:fitting}, but in what follows we prefer the inverse
fit (a linear fit that minimizes residuals in velocity width), which
best avoids slope bias.

\begin{deluxetable}{rrrrr}
\tablenum{\spirtfr}
\tabletypesize{\footnotesize}
\tablewidth{0pt}
\tablecaption{Tully-Fisher Fits to the Sa-Sd M$_{\rm R}^i<-18$ Sample~\tablenotemark{a}}
\tablehead{ & & RC Results & & H\,{\scriptsize I} Results \\
\cline{2-4} \cline{5-5}
\\
\colhead{Band} &
\colhead{Wtd Inv} &
\colhead{Bivariate} &
\colhead{Unwtd Inv} &
\colhead{Unwtd Inv} }
\startdata
\cutinhead{Slope} \\
U &  -9.54$\pm$0.29 &  -8.36$\pm$0.34 & -10.85$\pm$0.46 & -10.53$\pm$0.33 \\
B &  -9.52$\pm$0.27 &  -8.68$\pm$0.33 & -10.09$\pm$0.39 & -10.44$\pm$0.32 \\
R &  -9.70$\pm$0.27 &  -9.15$\pm$0.31 & -10.14$\pm$0.37 &  -9.75$\pm$0.27 \\
\cutinhead{Zero Point} \\
U & -19.60$\pm$0.04 & -19.71$\pm$0.05 & -19.82$\pm$0.06 & -19.71$\pm$0.04 \\
B & -19.71$\pm$0.04 & -19.79$\pm$0.05 & -19.83$\pm$0.05 & -19.74$\pm$0.04 \\
R & -20.69$\pm$0.04 & -20.76$\pm$0.05 & -20.81$\pm$0.05 & -20.62$\pm$0.03 \\
\cutinhead{Scatter~\tablenotemark{b}} \\
U & 0.88(0.56) & 0.81(0.51) & 1.02(0.63) & 0.85(0.45) \\
B & 0.78(0.56) & 0.73(0.52) & 0.82(0.59) & 0.77(0.46) \\
R & 0.70(0.56) & 0.67(0.53) & 0.74(0.58) & 0.68(0.42) \\
\enddata
\tablenotetext{a}{Fit results from weighted inverse, bivariate, and
unweighted inverse fitting techniques for optical RC (W$_{V_{pmm}}$) and
radio (W$_{\rm HI}$) T-F calibrations (see Appendix~\ref{sc:fitting}).  The
functional form of the TFR is M$_{\lambda}^i={ zero\, point}+
slope(\log{({\rm W}^i)} - 2.5)$.  Errors given are the formal statistical
errors from a two-step fit, see Appendix~\ref{sc:fitting}.  We require $i>40$.
The samples used for the different optical RC fits vary slightly, in that
one galaxy has no U-band data and another galaxy lies off the fitted
relation by $>$3$\sigma$ (the point rejection threshold) for some fits but
not others.}
\tablenotetext{b}{Measured biweight scatter and predicted scatter
(in parentheses) from measurement errors.}
\end{deluxetable}

Measurement-error scatter remains fairly constant across the three
passbands, while observed scatter increases sharply from R to B to U.
Subtracting the two in quadrature yields an estimate of the intrinsic
scatter in the TFR.  Using the 46 galaxies in the H\,{\small I}
linewidth sample, we have directly compared the intrinsic scatter
estimates from the H\,{\small I} linewidth and optical RC TFRs.  We
find a slightly lower intrinsic scatter for the optical TFR, which
probably indicates that the errors on the optical velocity widths have
been overestimated, and those on the catalog H\,{\small I} linewidths
underestimated (the H\,{\small I} errors do not account for systematic
effects such as confusion in the beam).  However, we have not adjusted
the predicted scatter values in Table~\spirtfr, as the listed values
would change by only $\sim$0.02 mag and the comparison performed here
is based on an incomplete subsample.

The steep slopes and high scatter values reported here may be surprising to
those familiar with restricted TF samples and/or alternate analysis
techniques based on forward fits (Appendix~\ref{sc:fitting}) and
luminosity-independent extinction corrections.  {\em We stress that the
slope, zero point, and scatter in different samples can only be
meaningfully compared when differences in sample selection and analysis
techniques are fully taken into account.}  In addition, it is important to
quantify measurement-error scatter, which may be qualitatively different
for cluster and field samples.  Below, \S\ref{sc:field}--\ref{sc:cluster}
compare the NFGS TF sample with several previous field and cluster TF
samples and demonstrate that our results are consistent when samples and
analysis techniques are carefully matched.  The reader who wishes to skip
directly to new results may go to \S\ref{sc:pecandsa}.

\subsection{Comparison with Field Galaxy Samples}
\label{sc:field}

The \citet{courteau:optical} and \citet[][hereafter
MFB]{mathewson.ford.ea:southern} samples are ideal for a direct comparison
with the NFGS because both employ optical rotation curves.  Both data sets
have been analyzed by \citet{courteau:optical} using the $V_{fit}$ velocity
width parameter described in Appendix~\ref{sc:vdefns}.  The Courteau sample
consists of $\sim$300 Sb--Sc galaxies with $55<i<75$, typically brighter
than $M_{R}^i\sim-18$ (using Courteau's extinction corrections) and pruned
by eye to eliminate peculiar and interacting galaxies
\citep{courteau:deep}.  The MFB sample includes $\sim$950 Sb--Sd galaxies
with $i>40$, typically brighter than $M_{R}^i\sim-18$ (assuming $R-I$
colors of $\sim$0.5 mag and using our standard extinction corrections).

\begin{figure}[tb]
\epsscale{0.8}
\plotone{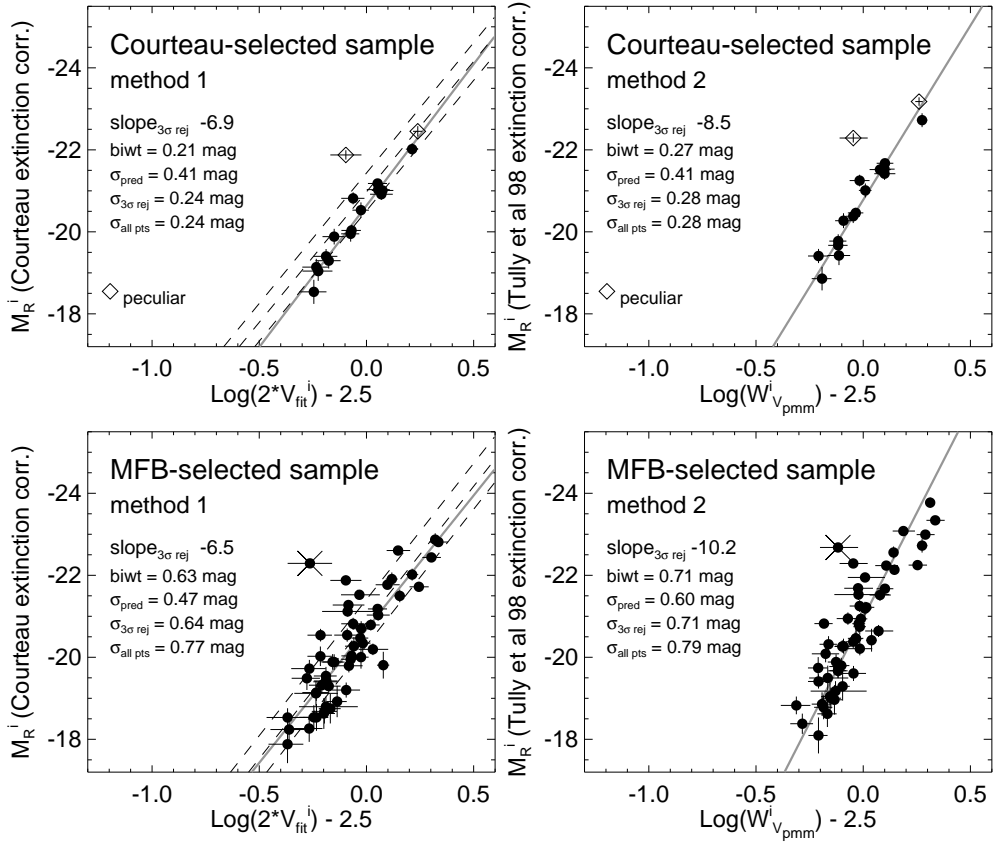}
\caption{Comparison between TF fits for the NFGS and for the samples of
\citet{courteau:optical} and \citet[][as analyzed by Courteau
1997]{mathewson.ford.ea:southern}.  The upper panels show NFGS data meeting
Courteau's selection criteria (interpreted to include Sab and Scd galaxies
to improve statistics), while the lower panels show NFGS data meeting MFB's
selection criteria (\S\ref{sc:field}).  Diamonds indicate
peculiar/interacting galaxies eliminated from the Courteau-selected sample
following Courteau (our peculiarity classification is discussed in
\S\ref{sc:pec}).  X's indicate a galaxy rejected from the MFB sample fits
as a $>$3$\sigma$ outlier (but still included in calculating the biweight
scatter and $\sigma_{\rm all\, pts}$).  {\em Left Panels:} NFGS comparison
samples analyzed using the methods of Courteau (forward fits,
luminosity-independent extinction corrections, $V_{fit}$ velocity
parameter, and q$_0=0.18$).  Printed parameters and thick gray lines give
fit results for the NFGS data.  Dashed lines show Courteau's forward fits
to the Courteau and MFB data sets and the $\pm1\sigma$ scatter on these
fits; both fits have been adjusted to H$_0=75$.  The Courteau fits have
also been shifted from Gunn r to Cousins R using r $-$ R = 0.354
\citep{jorgensen:secondary}, and the MFB fits have been shifted from
Cousins I to Cousins R using R $-$ I = 0.5 \citep[][their I$_c$ and
R$_c$]{frei.gunn:generating}.  {\em Right Panels:} The same data analyzed
using our standard techniques (unweighted inverse fits,
luminosity-dependent extinction corrections, W$_{V_{pmm}}$ velocity
parameter, and q$_0=0.20$).  Printed parameters and thick gray lines give
fit results for the NFGS data.}
\label{fg:cmath}
\end{figure}

Figure~\ref{fg:cmath} plots the TFRs for two subsamples of the NFGS defined
to match the Courteau and MFB selection criteria.  Except for the exclusion
of Sa--Sab galaxies, the Sb--Sd sample defined to match the MFB selection
criteria is identical to the spiral sample analyzed throughout
\S\ref{sc:spiralcal}--\ref{sc:spiral3}.  The left panels show Courteau's TF
fits for the MFB and Courteau samples overlaid on our matching subsample
data points and fits.  Here we use Courteau's analysis techniques,
including forward fits (minimizing residuals in M$_{\rm R}^i$), the
$V_{fit}$ velocity width parameter, and Courteau's luminosity-independent
extinction corrections.  For the 15 NFGS galaxies matching the Courteau
selection criteria, the biweight scatter of 0.21 mag is lower than
Courteau's 0.46, probably fortuitously.  The apparent zero point offset may
also represent small-number statistics, especially since no significant
zero point shift is apparent between the NFGS and MFB
samples.\footnote{However, a problem with the Courteau zero point may have
been seen elsewhere.  \citet{barton.geller.ea:tully-fisher} report that
their sample of galaxies in close pairs shows no zero point offset from the
Courteau sample but does show an offset of $\sim$0.4--0.5 mag from
\citet{tully.pierce:distances}, in the sense that the pairs sample is
brighter than the reference sample.  Given the likelihood that many of the
Barton et al.\ galaxies have experienced some luminosity enhancement due to
interactions, the result measured with respect to Tully \& Pierce seems
more likely to be correct, in which case the Courteau zero point would have
to be too bright (as we may be seeing here).}  For the 54 NFGS galaxies
matching the MFB selection criteria, the biweight scatter of 0.63 mag is
quite close to the value Courteau measures from the MFB data, 0.56.  The
slightly lower MFB scatter most likely reflects smaller measurement errors;
we return to this point below.

The right panels of Figure~\ref{fg:cmath} demonstrate the effect of
switching to our standard analysis techniques.  We use unweighted
inverse fits (minimizing residuals in velocity width), the
W$_{V_{pmm}}$ velocity parameter, and luminosity-dependent
extinction corrections \citep{tully.pierce.ea:global}.  With these
conventions, the slope of the TFR steepens considerably, and the
measured scatter about the fit increases, despite the smaller scatter
perceived by eye.  The eye perceives the scatter in velocity width,
whereas we measure the scatter in absolute magnitudes: the scatter in
absolute magnitudes increases as a result of the steeper slope of the
TFR.  This steeper slope is due in roughly equal measure to the use of
inverse fits and to the use of luminosity-dependent extinction
corrections.

Another factor affecting scatter is measurement errors.  In fact, the good
agreement between our scatter and the scatter in the MFB sample should
really be taken as evidence that the measurement errors in the two data
sets are very similar.  For the NFGS subsample defined by MFB's criteria
and analyzed using our standard techniques (lower right panel of
Figure~\ref{fg:cmath}), we calculate that 0.60 mag of the observed 0.71 mag
of scatter arises from measurement errors, leaving only $\sim$0.38 mag of
intrinsic scatter after subtraction in quadrature.  The NFGS subsample
error budget includes four contributions added in quadrature: photometry
errors and Hubble-law distance uncertainties (0.18 mag, dominated by the
contribution of the peculiar velocity field); photometric inclination
errors (0.42 mag, dominated by the one-decimal place precision of the UGC
axial ratios); velocity width uncertainties (0.38 mag, dominated by
intrinsic scatter in the correspondence between ``true'' rotation
velocities and measured optical velocity widths, regardless of high-quality
rotation curves); and extinction correction errors (0.09 mag, derived from
the uncertainties in inclination and velocity width).  For the MFB sample,
inclination errors should be somewhat smaller.  MFB galaxies are generally
larger on the sky than NFGS galaxies, so assuming that the photometric
inclinations for the MFB sample were computed using axial ratios with
round-off precision at least comparable to that of the UGC (and with at
least comparable accuracy), the MFB inclinations will almost certainly be
better than ours.  If we ascribe the (slight) scatter difference between
the MFB and NFGS samples entirely to the difference in inclination errors,
then we infer MFB inclination errors of $\sim$0.3 mag.

\clearpage

\subsection{Comparison with Cluster Samples}
\label{sc:cluster}

Except for its restriction to a single environment, the Ursa Major
sample of \citet[][hereafter VS]{verheijen.sancisi:ursa} is almost as
broadly representative as the NFGS.  Like the NFGS, the VS sample is
morphology-blind, including both Sa galaxies and peculiar and
interacting galaxies.  Here we restrict both samples to M$_{\rm
R}^i<-18$, type Sa--Sd, and $i>45$.  Within these limits VS's
H\,{\small I} data set is nearly complete, with only one Sa galaxy
missing.

\begin{figure}[tb]
\epsscale{0.5}
\plotone{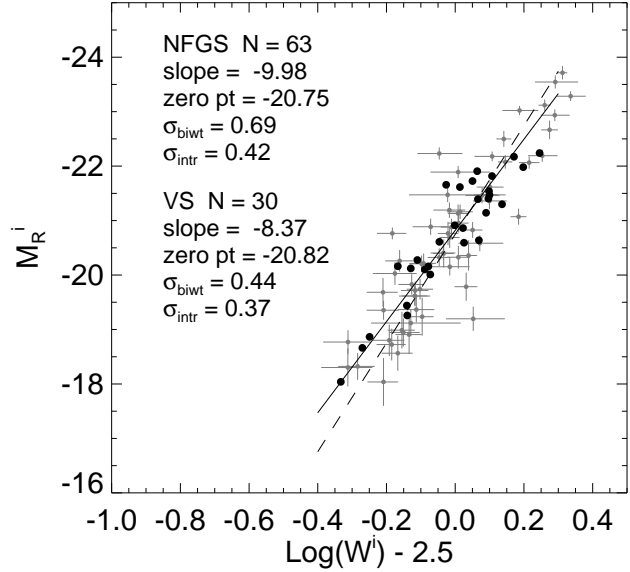}
\caption{Comparison of TFRs for the NFGS (gray) and for the Ursa Major
sample of \citet[][black]{verheijen.sancisi:ursa}, with both samples
restricted to types Sa--Sd, M$_{\rm R}<-18$, and $i>45$.  Solid and
dashed lines are unweighted inverse fits to the VS and NFGS data
respectively.  Biweight scatter values are with respect to the
corresponding fits.  Intrinsic scatter values are computed by
subtracting measurement-error scatter in quadrature (see
\S\ref{sc:cluster}).  We have recomputed VS's extinction corrections
using the original $W_{R}^i$ definition of
\citet{tully.fouque:extragalactic} to exactly reproduce the method
used for the NFGS corrections (\S\ref{sc:extinct}).  $W^i$ is
W$_{V_{pmm}}^i$ for the NFGS (\S\ref{sc:vcalcs}) and W$_{50}^i$ for
the VS sample; these linewidth measures are calibrated to the same
scale in Appendix~\ref{sc:vdefns}.  To match VS's distance scale (18.6
Mpc to Ursa Major), we temporarily adopt H$_0=77$ for NFGS galaxy
distances; this value of H$_0$ and the corresponding Ursa Major
distance were both derived from the same Cepheid-based distance
calibration \citep[][]{tully.pierce:distances}.  Note that we find an
Ursa Major distance of 16.9 Mpc, which would shift the cluster TFR
$\sim$0.2 mag fainter, if we simply apply the Tully \& Pierce value of
H$_0$ to the group velocity of Ursa Major ($V\sim1000$ km sec${-1}$
with respect to the Local Group) and correct for Virgocentric infall
at the coordinate center of Ursa Major according to the same
prescription used for NFGS recession velocities.}
\label{fg:verhfigscatt}
\end{figure}

Figure~\ref{fg:verhfigscatt} directly compares the NFGS and VS spiral
samples.  With distance scales, extinction corrections, and velocity width
definitions matched as described in the figure caption, we detect no zero
point offset within the errors.  The slope of
the VS sample appears considerably shallower, but inspection of the data
reveals that the behavior of a few galaxies at the faint end of the TFR
accounts entirely for the difference.  In \S\ref{sc:umdwarfs} we discuss
possible reasons for the divergence of the two samples at the faint end.

The sources of measurement error for the Ursa Major sample are
qualitatively different than for the NFGS, so we must compare intrinsic
scatter values, formed by subtracting measurement-error scatter from
observed scatter in quadrature.  We predict the measurement-error scatter
for VS using the same code with which we propagate our own errors, which
yields a total predicted R-band scatter of 0.24 mag from four sources:
hybrid kinematic-photometric inclination errors (0.09 mag); H\,{\small I}
linewidth errors (0.13 mag); distance uncertainties from cluster depth
effects \citep[0.17 mag, as estimated by][]{verheijen:ursa*1}; and errors
in photometry and extinction corrections (0.06 mag).  These numbers are
based on the inclination and H\,{\small I} linewidth errors given by VS
combined with basic photometry errors of $\sim$0.05 mag
\citep{verheijen:ursa*1}.  Because the NFGS error budget is dominated by
inclination errors, it is worth mentioning that VS's inclination errors are
smaller than ours not so much because they sometimes use kinematic
inclinations, as because their photometric inclinations are better than
ours.  Our inclination errors would probably be comparable to theirs if our
galaxies were equally large on the sky: the one-decimal-place precision of
the UGC adds greater round-off error for galaxies of smaller angular
diameter. To illustrate this, we have recomputed photometric inclinations
for the Ursa Major sample using UGC axial ratios and the method of
\S\ref{sc:vcorrs}.  Substituting these inclinations for VS's preferred
inclinations only slightly increases TF scatter.

After subtracting measurement-error scatter in quadrature, we obtain very
similar intrinsic R-band scatter values for the NFGS and the Ursa Major
sample: 0.42 mag and 0.37 mag.  Similar agreement is found in the B band.
Recomputing the NFGS scatter and errors about the shallower slope of the
Ursa Major sample, we find even better agreement, but if the NFGS velocity
width errors are slightly overestimated (\S\ref{sc:basiccal}) then the
intrinsic NFGS scatter about the Ursa Major slope may still be as high as
$\sim$0.42 mag.  The moderately higher intrinsic scatter of the NFGS
compared to Ursa Major should not be at all surprising, as the NFGS
includes a wide range of environments and star formation histories.  Apart
from the hint of a systematic offset at the faint end mentioned previously,
the two samples are in excellent agreement.

Other cluster TF studies quote lower scatter than we find for the complete
Ursa Major sample.  

\clearpage

\noindent For example, combining data from multiple clusters
(including Ursa Major, Fornax, Coma, the Pisces Filament, and several
others), \citet{tully.pierce:distances} measure an R-band scatter of 0.34
mag.  \citet{sakai.mould.ea:hubble} obtain the same result with a partially
overlapping data set and similar analysis techniques.\footnote{Sakai et
al.'s estimates of the intrinsic scatter of the TFR using the local
calibrator galaxies with Cepheid distances is not relevant to the present
discussion, as the calibrator galaxies do not even approximate a complete
sample of the local galaxy population.}  However, directly comparing the
NFGS spiral sample with these multi-cluster samples would be misleading,
for several reasons.  First, each individual cluster in the multi-cluster
samples obeys a different definition of ``complete'' and satisfies that
definition to a different degree.  For example, the ``complete'' Fornax
sample rejects interacting or disturbed galaxies and multiple systems
\citep{bureau.mould.ea:new}.  Second, each cluster represents a different
environment with a different star formation history, so each cluster's
color--magnitude relation may have a different color zero point
\citep[e.g.\ Tully \& Pierce find $B-I$ color zero point offsets of
$\sim$0.1 mag when they compare cluster color--magnitude relations; see
also][]{watanabe.yasuda.ea:surface}.  We will see in \S\ref{sc:color} that
if real, such color offsets almost certainly imply TF zero point offsets
between the clusters.  However, the TF scatter due to these TF zero point
offsets is suppressed in the multi-cluster analyses to the extent that the
individual cluster TFRs are allowed to shift freely to minimize zero point
offsets (the absolute zero point is generally determined separately using
the Cepheid calibrator galaxies).  A third concern is the effect of a
top-heavy luminosity distribution: in combining more distant clusters with
more nearby ones, the multi-cluster studies define samples that
statistically favor bright galaxies.  This bias may drive down scatter if
fainter galaxies have higher scatter.  While the existence of such a trend
is unclear in our restricted spiral sample, dwarf galaxies definitely show
higher scatter in the full sample (\S\ref{sc:fulltfr}).

\section{The Spiral TFR 2:  Sa \& Peculiar Galaxies}
\label{sc:pecandsa}

We have shown that our TF results are consistent with previous studies if
we reproduce their sample selection criteria.  These criteria often exclude
Sa galaxies and/or galaxies with morphological peculiarities.  Here, we
begin the process of analyzing TF scatter for a broad spiral sample by
turning the spotlight on Sa and peculiar galaxies, highlighting where these
galaxies' TF residuals fall within the general spiral TFR.  For this
analysis we return to the full NFGS spiral TF sample: types Sa--Sd brighter
than M$_{\rm R}^i=-18$ and inclined by $>$40$\degr$.

\subsection{An Sa Galaxy Offset}
\label{sc:saoffset}

\begin{figure}[tb]
\epsscale{1.}
\plotone{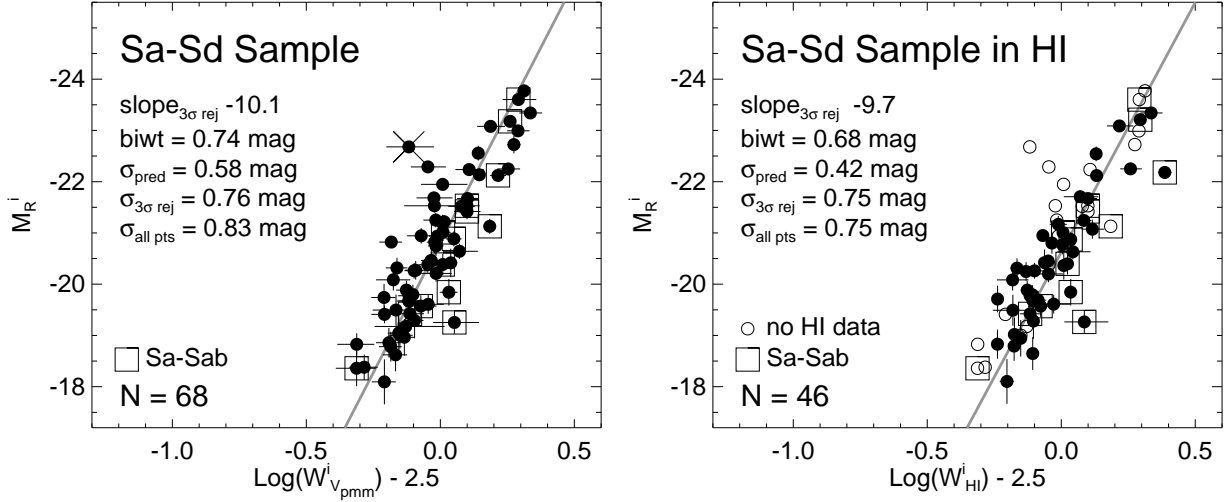}
\caption{TFR for the Sa--Sd sample, illustrating the Sa galaxy offset
toward higher $W^i$/lower $L$.  The left panel uses optical velocity
widths while the right panel uses H\,{\small I} linewidths, with
missing data points indicated by an open circle at the position of the
optical velocity width.  Lines show unweighted inverse fits to the
solid points.}
\label{fg:sa}
\end{figure}

Fourteen of the 68 galaxies in the NFGS spiral TF sample are Sa--Sab
galaxies, collectively referred to as ``Sa'' galaxies from here on.  As a
group, these 14 galaxies sit clearly to one side of the TFR
(Figure~\ref{fg:sa}), with offsets toward lower $L$/higher $W^i$ of
0.76 mag at R, 0.95 mag at B, and 1.20 mag at U as compared to Sb--Sd
galaxies (where both Sa and Sb--Sd offsets are computed with respect to the
unweighted inverse TF fits in Table~1, with relative offset uncertainties
of $\sim$0.17 mag).

The possibility that Sa galaxies rotate faster at a given luminosity was
first suggested by \citet{roberts:twenty-one} based on 21 cm data.  Sa
offsets became controversial after \citet{rubin.burstein.ea:rotation}
claimed that the Sa TFR lies 2 mag below the Sc TFR in the B band.  Other
studies reported smaller offsets \citep[e.g.][]{pierce.tully:distances}.
\citet{simard.pritchet:internal} argue that a greater intrinsic spread of
luminosities for late-type vs.\ early-type galaxies at a given velocity
width, combined with Malmquist bias, may explain the large offset found by
\citet{rubin.burstein.ea:rotation}.  However, Sa offsets continue to be
observed in increasingly complete samples \citep[e.g.][see
\S\ref{sc:color}]{verheijen:ursa}.

The scatter properties of our data suggest that the offset we observe
is mostly real.  In the R band, the Sa galaxies and the Sb--Sd
galaxies have biweight TF scatter values of 0.63 mag and 0.70 mag,
respectively.  These numbers imply that Malmquist bias is unlikely to
produce the observed 0.76 mag offset between the two populations.  In
the B and U bands, the scatter in each population increases but the
offset increases even more.  The offset exceeds the scatter by 0.3 mag
at U.

We have investigated whether the Sa offset in our data may be ascribed
to systematic errors in luminosity or velocity width measurements.
For example, large bulges could make Sa galaxies appear rounder (and
more face on) than they really are, leading to underestimated
extinction corrections and overestimated $\sin i$ corrections.  Sa's
in the present sample do have more face-on inclination estimates on
average (61$\degr$ vs. 66$\degr$ for the full spiral TF sample);
however, this difference can account for an offset of only $\sim$0.2
mag at a TF slope of $-$10.  Furthermore, even extreme assumptions
about axial ratios (e.g.\ assigning an intrinsic axial ratio of
$q_0=0.4$ to Sa's and assigning $q_0=0.09$ to all other types) reduce
the R-band offset by at most half and have much less effect in the B
and U bands.  No other aspect of our analysis is responsible for the
offset either.  Whether we adopt luminosity-dependent or independent
extinction corrections has negligible effect on our results.  We do
not add an explicit morphological type dependence to our extinction
corrections, but in any case adding such a dependence would probably
only increase the Sa offset: extinction is likely to be lower in Sa
galaxies \citep{kodaira.watanabe:statistical}.  The offset is not
appreciably affected when we substitute $V_{fit}$ for $V_{pmm}$ or
when we omit the optical-to-radio conversion.  Finally, the available
H\,{\small I} data also seem to confirm the offset
(Figure~\ref{fg:sa}), although these data are subject to
incompleteness and high scatter for the Sa population.

We conclude that the bulk of the Sa offset in our sample is real. However,
with just 14 Sa galaxies, our analysis is subject to small number
statistics.  Formally, a Kolmogorov-Smirnov test yields a probability of
$6\times10^{-4}$ that the R-band TF residuals of the Sa and Sb--Sd
subpopulations were drawn from the same parent population
($\sim$3.5$\sigma$).  A possible physical explanation for the Sa offset and
a way to remove the offset to reduce scatter will be discussed in
\S\ref{sc:color}.

\subsection{Galaxy Peculiarity and Sample Pruning}
\label{sc:pec}

\begin{figure}[tb]
\epsscale{0.5}
\plotone{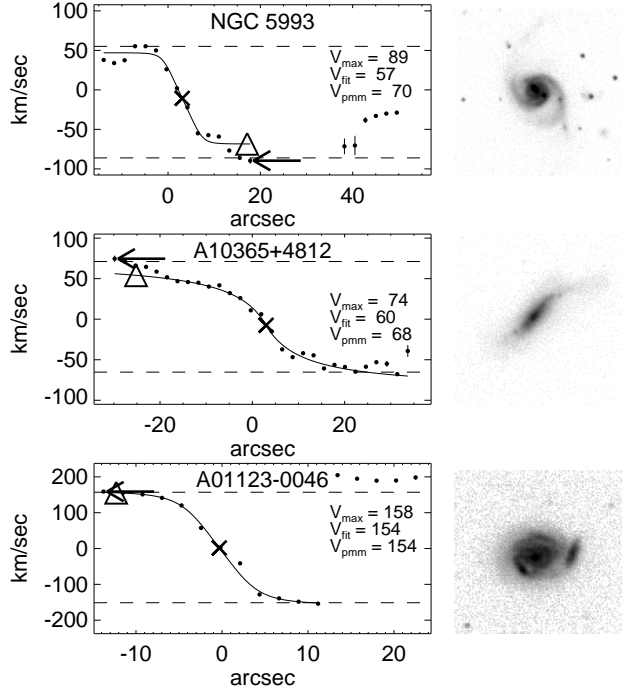}
\caption{B-band images and rotation curves for three galaxies with
morphological peculiarities (\S\ref{sc:pec}).  Notation is as in
Figure~\ref{fg:rcgallery}.  The distinct kinematic components at large
radii in the first and third panels were not used in computing velocity
widths.  Images are courtesy of \citet{jansen.franx.ea:surface}.}
\label{fg:pecgallery}
\end{figure}

TF samples are frequently pruned to remove morphologically peculiar
galaxies and sometimes also to remove barred galaxies.  Here we use the
term ``peculiar galaxies'' to mean disturbed spiral galaxies (i.e.\ those
with warps, tidal features, multiple nuclei, polar rings, or interacting
neighbors) rather than galaxies that cannot be reliably classified as
spirals.  Figure~\ref{fg:pecgallery} shows rotation curves and images for
three NFGS galaxies we have identified as morphologically peculiar.  To
avoid bias, bars and morphological peculiarities were independently
identified by two of us (S.K. and M.F.) without reference to TF residuals,
and the identifications were further checked against the notes provided by
R.A. Jansen in \citet{jansen.franx.ea:surface}.  Although a number of weak
bars and peculiarities were noted, we report here only strong, unambiguous
cases identified by more than one observer.  Such cases are broadly
distributed across a wide range of spiral types and luminosities.

\begin{figure}[tb]
\epsscale{0.8}
\plotone{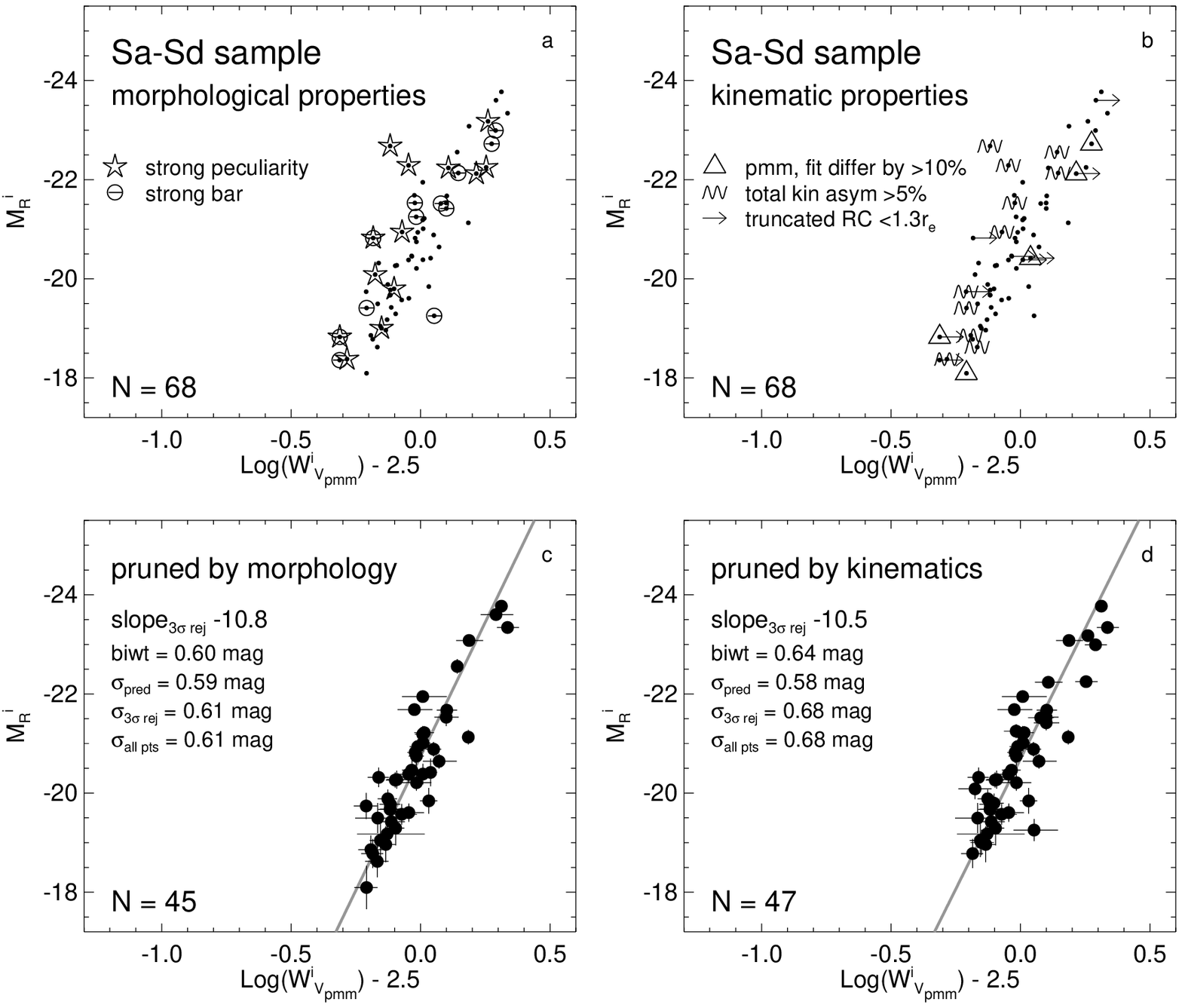}
\caption{Upper Panels: Location of kinematically and morphologically peculiar
galaxies within the Sa--Sd TFR.  Lower Panels: Effect of pruning the sample
to remove such galaxies.  Unweighted inverse fit results are shown.}
\label{fg:mfbprune}
\end{figure}

Figure~\ref{fg:mfbprune}a uses two different (possibly overlapping) symbols
to identify peculiar and barred galaxies within the NFGS TFR.  Although the
barred galaxies that are {\em not} peculiar do not show a clear systematic
offset, the peculiar galaxies as a group tend to lie on the high $L$/low
$W^i$ side of the TFR.  This pattern suggests that pruning such galaxies
may obscure interesting physics hidden in TF scatter.

Even when the goal is scatter reduction, pruning may not be the most
effective strategy.  Figure~\ref{fg:mfbprune}c demonstrates that pruning
barred and peculiar galaxies from the NFGS spiral sample reduces R-band TF
scatter to approximately the level expected from measurement errors.  In
bluer passbands, however, the reduced scatter still exceeds the level
expected from measurement errors: we measure 0.63 mag and 0.71 mag in B and
U, respectively, compared to a predicted measurement-error scatter of 0.58
mag in both passbands.  Also, the decrease in scatter won by pruning comes
at a high price, as $>$30\% of the sample must be rejected.  In
\S\ref{sc:spiral3} we will use more objective, physically motivated
techniques to reduce TF scatter to a level comparable to measurement-error
scatter in all passbands without discarding galaxies from the sample.

We have also considered independent definitions of peculiarity involving
kinematic data: Figure~\ref{fg:mfbprune}b identifies galaxies with
truncated rotation curves ($r_{max}<1.3r_e$, \S\ref{sc:vcalcs}), rotation
curve asymmetries greater than 5\% (our RC asymmetry index is defined in
Appendix~\ref{sc:asymmeast}), or discrepancies between W$_{V_{fit}}$ and
W$_{V_{pmm}}$ greater than 10\%.  Interestingly, galaxies with high RC
asymmetry behave much like those with strong morphological peculiarities,
with high $L$/low $W^i$ offsets.  Figures~\ref{fg:mfbprune}c
and~\ref{fg:mfbprune}d show that pruning based on either morphology or
kinematics produces a comparable reduction in scatter, although kinematic
pruning misses some Sa outliers.

To test whether our definitions of kinematic peculiarity have physical
significance or merely identify faulty rotation curves, we have tried
pruning the H\,{\small I} linewidth TFR using kinematic peculiarity
information obtained from our optical RCs.  Amazingly, optical RC
peculiarities are at least as effective as morphological peculiarities in
flagging outliers in the H\,{\small I} linewidth TFR.  This result implies
that either (1) optical RC peculiarities are associated with observational
errors in the NFGS surface photometry or the UGC-derived photometric
inclinations common to both the H\,{\small I} and optical RC TFRs, or (2)
optical RC peculiarities occur in galaxies that lie off the TFR for
physical reasons.  We find evidence for the latter in \S\ref{sc:spiral3}.

\section{The Spiral TFR 3:  Third Parameters \& Physical Sources of Scatter}
\label{sc:spiral3}

\begin{figure}[tb]
\epsscale{0.5}
\plotone{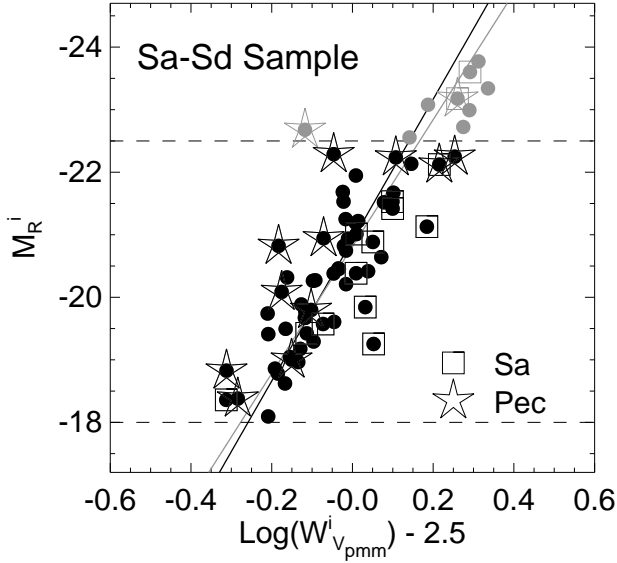}
\caption{Systematic offsets in the Sa--Sd TFR.  Sa and peculiar
galaxies fall on opposite sides of the TFR, helping to drive strong
correlations between TF residuals and physical properties such as
effective $B-R$ color and global EW(H$\alpha$).  In \S\ref{sc:spiral3}
we quantify the statistical significance of these correlations within
the luminosity range between the dashed lines.  This range excludes
the asymmetric scatter at the bright end of the relation to ensure
fair statistical tests (\S\ref{sc:thirdpartech}).  The gray and black
lines show unweighted inverse TF fits to the sample with and without
galaxies brighter than M$_{\rm R}^i=-22.5$.  For statistical tests, we
define residuals relative to the black line.}
\label{fg:tfr}
\end{figure}

We have seen that interacting, merging, and peculiar galaxies often
lie on the high $L$/low $W^i$ side of the TFR, while Sa galaxies lie
on the low $L$/high $W^i$ side
(\S\ref{sc:pecandsa}). Figure~\ref{fg:tfr} shows that in addition,
NFGS galaxies brighter than M$_{\rm R}^i\sim-22.5$ display asymmetric
scatter about the best-fit TFR for fainter galaxies, with a systematic
offset of $\sim$0.7 mag toward lower $L$/higher $W^i$.  Below we
demonstrate that all of these offsets reflect underlying correlations
between TF residuals and quantitative galaxy properties, most notably
$B-R$ color and H$\alpha$ equivalent width (EW).  Following tradition,
we refer to these properties as ``third parameters,'' i.e.\ additional
variables that control physical scatter in the two-parameter TFR.

\subsection{Third Parameter Analysis Technique}
\label{sc:thirdpartech}

\citet{barton.geller.ea:tully-fisher} demonstrate that false
correlations between TF residuals and candidate third parameters may
arise when the TF slope is measured incorrectly, in which case any
parameter that varies along rather than perpendicularly to the TFR
will produce a correlation.  A true third parameter should vary at
least partly perpendicularly to the TFR.  

We adopt the following strategy to avoid false detections of third
parameters.  First, we apply strict cuts in M$^{i}$ to eliminate
sections of the data where the velocity width scatter is clearly
asymmetric.  For the R band, these cuts are indicated by the dashed
lines in Figure~\ref{fg:tfr} (see also Table~\lumcuts): the upper line
excludes galaxies brighter than M$_{\rm R}^i\sim-22.5$, which show a
systematic offset as discussed above, and the lower line excludes
galaxies that do not meet the M$_{\rm R}^i<-18$ sample definition
criterion.  After these cuts, the data are biased in M$^{i}$ but
nearly perfectly unbiased in velocity width.  Second, we apply an
inverse TF fit to the data, which minimizes residuals in velocity
width.  At this point the velocity width residuals are completely
uncorrelated with luminosity M$^{i}$.  Finally, we use Spearman rank
tests to quantify both the candidate third parameter--TF residual
correlation and the candidate third parameter--luminosity correlation.
For the Spearman tests, velocity width residuals and magnitude
residuals are interchangeable.  If these tests show that the candidate
third parameter correlates with TF residuals but not with luminosity,
then we conclude that it varies perpendicularly to the TFR.

\begin{deluxetable}{llll}
\tablenum{\lumcuts}
\tabletypesize{\footnotesize}
\tablewidth{0pt}
\tablecaption{Magnitude Cuts for Spearman Rank Tests~\tablenotemark{a}}
\tablehead{\colhead{} &
\colhead{U} &
\colhead{B} &
\colhead{R} }
\startdata
bright cut		& -21 & -21 & -22.5 \\
standard faint cut	& -17 & -17 & -18   \\
dwarf faint cut		& -16 & -16 & -17   \\
\enddata
\tablenotetext{a}{Cuts applied to extinction-corrected absolute
magnitudes for third-parameter tests.  See \S\ref{sc:thirdpartech}.}
\end{deluxetable}

\begin{figure}[tb]
\epsscale{1.}
\plotone{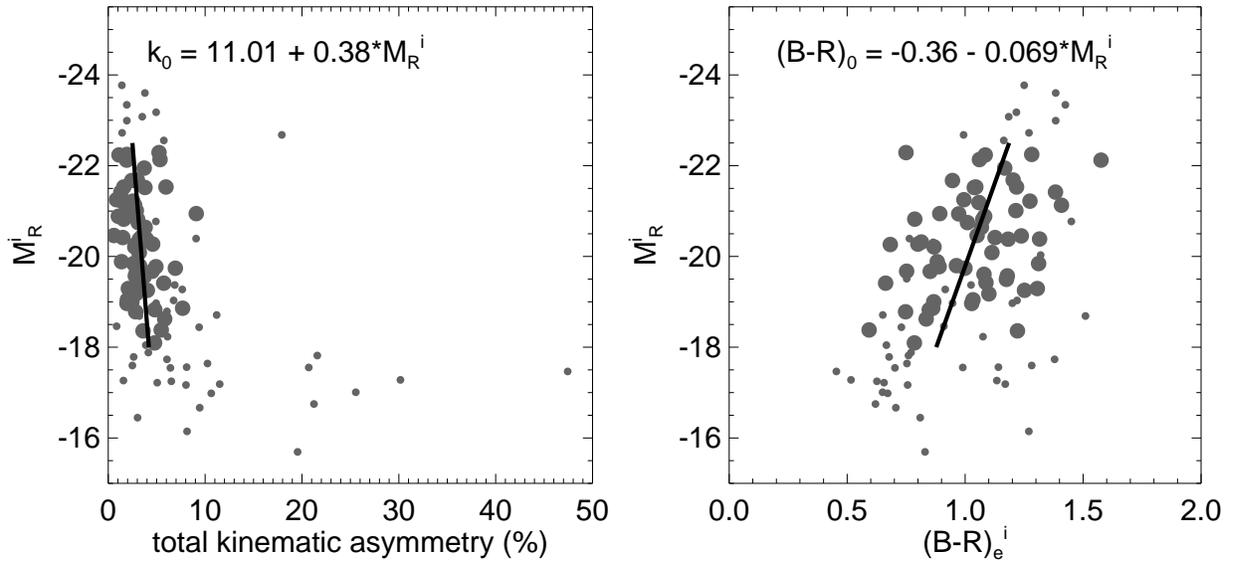}
\caption{M$_{\rm R}^i$ vs.\ total RC asymmetry
(Appendix~\ref{sc:asymmeast}) and effective $B-R$ color.  Larger points
represent the luminosity and inclination-restricted spiral sample of
Figure~\ref{fg:tfr}, while smaller points show other galaxies in the
optical RC sample.  Galaxies with $i<40$ are excluded from both panels.
Black lines represent least-squares inverse fits to the larger points,
minimizing residuals in color and RC asymmetry.  Note that inverse fits are
appropriate for combining with our inverse TF fits; some bias is clear in
the line slopes, but it has no effect on our conclusions.}
\label{fg:kandcvslum}
\end{figure}

In fact, even if a parameter does vary with luminosity, it may still
qualify as a third parameter if it also correlates clearly with TF
residuals, because we have in principle eliminated the TF
residual--luminosity correlation.  Nonetheless, one must be wary if
the parameter-luminosity correlation is stronger than the
parameter--TF residual correlation, as in such a case even small TF
slope errors may lead to false correlations.  For luminosity-dependent
parameters, one may wish to subtract out the luminosity dependence and
measure the correlation between TF residuals and ``parameter
residuals,'' i.e.\ parameter offsets from the fitted
parameter--luminosity correlation \citep[cf.][]{courteau.rix:maximal}.
Below we give results based on raw parameters, but we have also
checked our results using parameter residuals.  The
parameter--luminosity correlations are strongest for color and
rotation curve asymmetry (Figure~\ref{fg:kandcvslum}); however, even
for these parameters we find that correcting for the luminosity
dependence makes little difference, slightly improving the
third-parameter correlation if anything.

In general, we use unweighted TF fits to define TF residuals.
Unweighted fits avoid the slope bias that would otherwise arise
because our errors vary systematically with luminosity and velocity
width.  However, we have confirmed all correlations using both
weighted and unweighted fits.  We have also confirmed all correlations
using two definitions of velocity width (W$_{V_{fit}}$ and
W$_{V_{pmm}}$) and using both luminosity-dependent and
luminosity-independent extinction corrections (\S\ref{sc:extinct}).
Results quoted in the text use W$_{V_{pmm}}$ and the
\citet{tully.pierce.ea:global} extinction corrections.

\clearpage

\subsection{Third Parameter Test Results}
\label{sc:thirdparres}

We have tested a wide range of candidate third parameters, making use of
photometric and spectrophotometric quantities from the NFGS database
\citep{jansen.fabricant.ea:spectrophotometry,jansen.franx.ea:surface} and
photometric asymmetries kindly provided by R. Jansen \citep[private
communication, see also][]{jansen:nearby}.  For the luminosity-restricted
Sa--Sd sample of Figure~\ref{fg:tfr}, TF residuals correlate strongly with
$(B-R)_e$ (effective $B-R$ color measured within the B-band half-light
radius) and global EW(H$\alpha$) in all bands.  Other measures of color and
emission-line strength yield somewhat noisier correlations.  We also find
correlations with rotation curve asymmetry, photometric asymmetry, and two
measures of surface brightness; these correlations reach $>$3$\sigma$
significance only in certain photometric bands.  TF residuals do not
correlate with luminosity (by construction) or with inclination, validating
our analysis.  Likewise, we detect no correlation between TF residuals and
rotation curve extent, isophotal radius, effective radius, H\,{\small I}
gas mass, gas consumption timescale, or $M_{\rm HI}/L$ (we have incomplete
data for the last three however, see \S\ref{sc:neutral}).

\begin{figure}[tb]
\epsscale{0.8}
\plotone{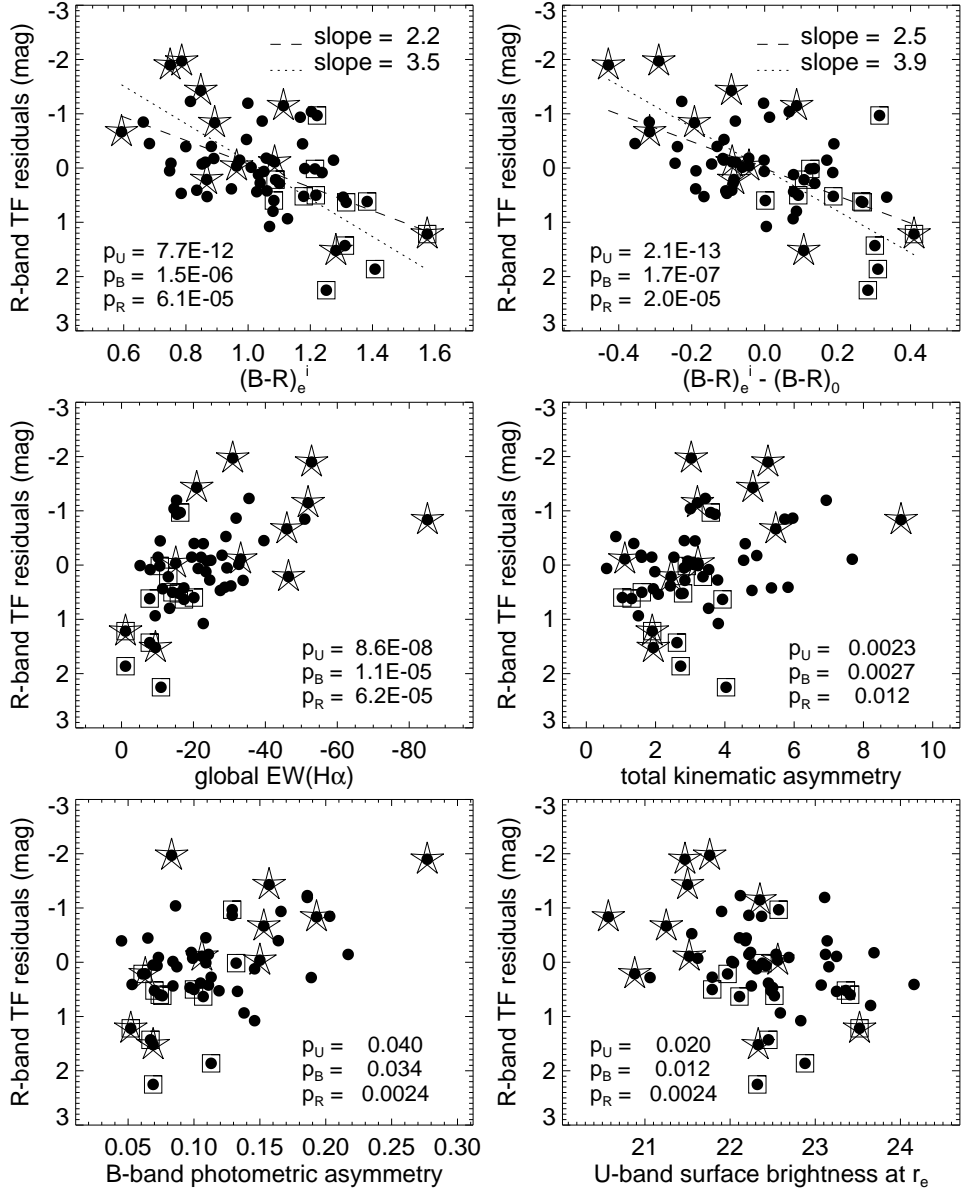}
\caption{TF residuals vs.\ physical properties of the Sa--Sd $i>40$
$-18<\rm M^i_{\rm R}<-22.5$ sample (see also Figure~\ref{fg:tfr}).  Squares
indicate Sa galaxies and stars indicate galaxies with morphological
peculiarities (\S\ref{sc:pec}).  The Spearman rank probability of no
correlation between the parameter and TF residuals in each optical band is
given as $p_{band}$.  Only R-band TF residuals are plotted.  Dashed and
dotted lines in the top panels show forward least-squares and least-squares
bisector fits \citep[][see \S\ref{sc:loffs}]{isobe.feigelson.ea:linear}.
Colors in the top right panel are defined relative to the color--magnitude
relation (Figure~\ref{fg:kandcvslum}).}
\label{fg:restspircorr}
\end{figure}

Figure~\ref{fg:restspircorr} displays some of the correlations in the R
band and lists Spearman rank test results for U, B, and R (the
``probability of no correlation'' $p$ is given, where $1-p$ is the
confidence of the result).  The symbol coding shows that Sa and
peculiar galaxies drive most of the statistical signal in the R-band
correlations.  If we eliminate both groups of galaxies, no R-band
correlation survives with $>$3$\sigma$ confidence.  However, the
$(B-R)_e$ correlation does survive in B, and both color and emission
line correlations remain significant in U.  In general, the two groups
of galaxies contribute roughly equally to the correlations, with the
Sa galaxies showing a slightly stronger signal due to lower scatter.
The peculiar galaxies form a more heterogeneous group, including not
only interacting, merging, and warped galaxies, but also a few
galaxies whose oddities may indicate later evolutionary states, e.g.\
a bulge-dominated Sa galaxy with a prominent polar dust lane
(NGC~984).

Although Sa and peculiar galaxies dominate the statistical signal we
measure, the correlations in Figure~\ref{fg:restspircorr} apply to all
galaxies in the sample.  A blue, strongly star-forming galaxy whose
image does not appear peculiar nonetheless shows the expected TF
residual.  Thus the correlations provide a physical, quantitative
basis for understanding systematic offsets in the TFR, independent of
morphological classification.

Furthermore, the correlations are not limited to the optical TFR.  Despite
disproportionate H\,{\small I} catalog incompleteness for bright and
peculiar galaxies, mixed H\,{\small I} data quality, and confusion in
single-dish linewidths, we still detect 2--3$\sigma$ correlations between
radio TF residuals and both $(B-R)_e$ and global EW(H$\alpha$).  We also
find evidence for a TF residual--total color correlation in radio data from
the complete Ursa Major database of \citet[][]{verheijen.sancisi:ursa}, see
\S\ref{sc:color}.  These results emphasize the close link between optical
and radio velocity widths already noted in \S\ref{sc:vcalcs} and
\S\ref{sc:pec} (see also Appendix~\ref{sc:vdefns}).  However, we do not
perform a detailed third-parameter analysis of the radio TFR, as the
catalog H\,{\small I} linewidths are incomplete and of mixed quality.

\clearpage

\subsubsection{Color}
\label{sc:color}

Effective $B-R$ color yields a striking correlation with TF residuals,
predicting spiral galaxy TF offsets better than any other physical
parameter we have tested.  Effective $U-B$ color also produces a clear
correlation in all three bands.  The color--TF residual correlations
strengthen from R to B to U, as blue peculiar galaxies and red Sa galaxies
move further apart in the TFR.  Although we correct colors for internal
extinction (see \S\ref{sc:extinct}), omitting the extinction correction
does not significantly affect the correlation strength.  The correlation
slope is quite a bit steeper than expected from dust extinction and
reddening \citep[$\sim$1.3 in the R band with $B-R$ colors,
e.g.][Table~1]{gordon.calzetti.ea:dust}.

We stress that the correlation reported here cannot be dismissed as merely
the ``built-in'' correlation due to the variation of TF scatter with
passband.  A weak correlation of U and B-band TF residuals with color might
have been expected from the fact that TF scatter is significantly higher in
these bands than in the R band, which implies that both $B-R$ and $U-R$
colors may be used to reduce scatter (most trivially, simply by adding
these colors to the U and B magnitudes to recover the R magnitudes).  In
this case, the color--TF residual correlation would have a slope of $\sim$1
and there would be no correlation between R-band TF residuals and $B-R$
colors.  However, both the persistence of the correlation in the R band and
the steep slope of the correlation point to a more interesting result: we
explore this result empirically in this section and discuss its physical
significance in \S\ref{sc:causes}.

For our sample, effective colors produce much tighter correlations
than total colors.  In the R band, Spearman tests yield
$p=6.1\times10^{-5}$ for the ($B-R$)$_e^i$ correlation, or
$\sim$4$\sigma$, but only $p=0.034$ for the total ($B-R$)$^i$
correlation, or just over 2$\sigma$. Total colors dilute the
correlation signal with observational scatter from profile
extrapolation and background subtraction (R. Jansen, private
communication).  However, we see no strong systematic differences
between $B-R$ and $(B-R)_e$: on average the two track closely, with
$(B-R)_e$ slightly redder, by at most $\sim$0.1 mag for the reddest
galaxies.

\begin{figure}[tb]
\epsscale{.5}
\plotone{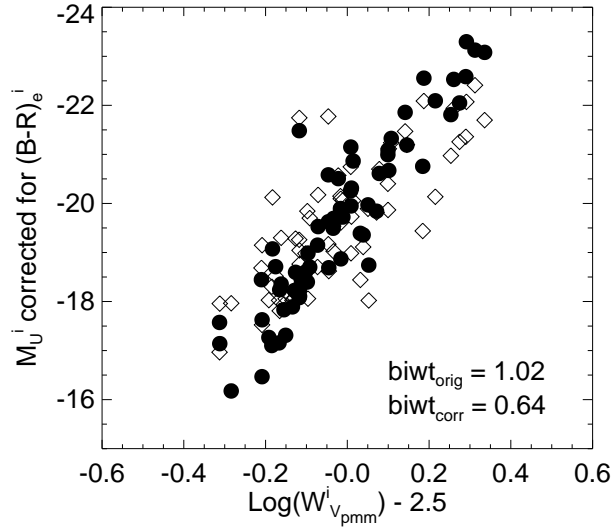}
\caption{Scatter reduction in the U-band TFR using a fit to the
effective color--TF residual correlation (\S\ref{sc:color}).  Diamonds show the
original points and dots show the color-corrected points.  Note that
the corrected points define a steeper slope due to the
color-magnitude relation.  The biweight scatter for the original and
color-corrected magnitudes is given in mag.}
\label{fg:scattred}
\end{figure}

The color--TF residual correlation offers an alternative to sample
pruning for reducing TF scatter.  If we restore spiral galaxies
brighter than M$_{\rm R}^i=-22.5$ to the sample (they were not used in
Figure~\ref{fg:restspircorr}, see Figure~\ref{fg:tfr}), then refitting
the color--TF residual correlation provides a color-correction
formula to go with the unweighted inverse TF fit in Table~\spirtfr.  A
least-squares forward fit to $\Delta$M vs.\ $(B-R)_e^i$ yields
\begin{equation}
\label{eq:colcorr}
\Delta \rm M_{\rm R}^i = -2.0 +1.8(B-R)_e^i
\end{equation}
where $\Delta \rm M_{\rm R}^i$ is the measured TF residual (observed magnitude minus fitted relation).
If we correct our R-band magnitudes with equation~\ref{eq:colcorr} and
recompute the residuals with respect to the original TF fit, the
biweight scatter falls from 0.74 mag to 0.66 mag.  An analogous
procedure reduces scatter to 0.64 mag in both the B and U bands, from
0.82 and 1.02 mag respectively, essentially eliminating the
differences between wavelength bands (Figure~\ref{fg:scattred}).  The
corrected scatter values come quite close to the scatter expected from
measurement errors (Table~\spirtfr).  Determining whether these errors hide
further non-random behavior in TF scatter would require smaller
measurement errors.  Full two-dimensional velocity fields, with
kinematic inclinations and modeling of disk warps and asymmetries,
offer the best hope of progress
\citep[cf.][]{bershady.andersen:evolution}.

The existence of a strong color--TF residual correlation is not entirely
unexpected.  \citet{pierce.tully:luminosity-line} describe environmental
shifts in the B-band TFR for a mixed cluster-field sample that are linked
to galaxy colors, which suggests the possibility of a color--TF residual
correlation.  \citet{giraud:two-color} demonstrates reduced TF scatter in a
two-color Tully-Fisher relation (essentially equivalent to our
color-corrected TFR, as in Figure~\ref{fg:scattred}, but with the slope of
the color--TF residual correlation assumed rather than measured).
\citet{rhee:physical} reduces TF scatter using 60 $\mu$m data to define
``population corrected'' luminosities.  \citet{bershady.haynes.ea:rotation}
present initial evidence for a color--TF residual correlation in their
intermediate-redshift sample, which spans a wide range of galaxy types;
however, the TF residuals are measured relative to an externally defined
TFR, making it harder to eliminate systematic errors (also a potential
problem for \citet{zaritsky:evidence}, see
\citet{barton.geller.ea:tully-fisher} and the discussion in
\S\ref{sc:thirdpartech}).  Most significantly, \citet{verheijen:ursa*1}
reports a $B-I$ color vs.\ TF residual correlation for a subsample of 21
galaxies drawn from his complete Ursa Major sample, and our own analysis of
his full Sa--Sd data set indicates that total $B-R$ colors correlate with
B-band TF residuals at $\sim$3--3.5$\sigma$ confidence \citep[using
unweighted inverse fits, our standard luminosity cuts, and the W$_{50}$ and
M$_{\rm B}^i$ from][]{verheijen.sancisi:ursa}.  The R-band correlation for
this sample registers at $\sim$2$\sigma$, consistent with our expectations
given that total rather than effective colors were used.

Most other low-redshift TF studies exclude Sa and peculiar galaxies
and/or use red to near-infrared photometry, making it difficult to
observe the color--TF residual correlation.  For example,
\citet{courteau.rix:maximal} find only a slight, statistically
insignificant color--TF residual correlation in a sample of
undisturbed Sb--Sc galaxies observed in the R band.  The close galaxy
pairs sample of \citet{barton.geller.ea:tully-fisher} includes
disturbed galaxies but was not designed to evenly sample the range of
spiral morphologies; again no correlation emerges using total colors
and R-band TF residuals (E. Barton, private communication).

The color--TF residual correlation provides a natural explanation for
two effects noted previously: the Sa galaxy offset
(\S\ref{sc:saoffset}) and the asymmetric scatter at the bright end of
the TFR (\S\ref{sc:spiral3}).  Sa galaxies are typically red and show
relatively little scatter in color compared to later type galaxies
\citep{jansen.franx.ea:surface}.  As a consequence, Sa galaxies
deviate from the TFR as a group, with larger offsets in bluer bands.
Applying the color correction given in Equation~\ref{eq:colcorr}
reduces the R-band offset between Sa and Sb--Sd galaxies from
0.76$\pm$0.17 to 0.28$\pm0.23$ mag, where we have computed offsets
using TF residuals relative to the unweighted inverse fit in Table~1.
The remaining offset is not statistically significant.  Similarly, the
$\sim$0.7 mag offset for galaxies brighter than M$_{\rm R}^i=-22.5$
shrinks to $\sim$0.15 mag following the color correction, suggesting
that the asymmetric scatter at the bright end of the TFR may simply
reflect the rarity of extremely luminous blue galaxies.

\citet{verheijen:ursa*1} has also reported a link between color dependence
and morphological type dependence in the TF residuals for his pruned Ursa
Major spiral sample.  The Sa and Sab galaxies are underluminous by
$\sim$0.5 mag in the B band (consistent with our result when normalized for
slope differences, $-$6.7 vs.\ $-$10), while the same galaxies have normal
luminosities at K, suggesting a color effect.  Later types show a wide
range of positive and negative TF residuals in both Verheijen's complete
TF sample \citep{verheijen:ursa} and in the NFGS spiral TF sample, consistent
with the much greater color dispersion observed for types Sb and later in
the full NFGS \citep{jansen.franx.ea:surface}.  We discuss a physical
interpretation of the color--TF residual correlation in terms of stellar
populations in \S\ref{sc:causes}.

\subsubsection{Emission Line Strength}
\label{sc:ewha}

Global H$\alpha$ equivalent width \citep[integrated over the entire
galaxy, see][]{jansen.fabricant.ea:spectrophotometry} correlates with
TF residuals for the Sa--Sd subsample almost as well as $(B-R)_e$
(Figure~\ref{fg:restspircorr}).  This result should not be too
surprising: the two parameters are highly correlated.  EW(H$\alpha$)
measures recent star formation on timescales of Myr relative to
integrated past star formation, while $(B-R)_e$ measures the ratio
between young and old stellar populations and thus traces star
formation on timescales of Gyr
\citep[e.g.][]{kennicutt.tamblyn.ea:past}.  As expected, the
EW(H$\alpha$)--TF residual correlation intensifies in bluer TF bands,
although not quite so strongly as the color correlation.  Substituting
either nuclear (fixed $3^{\prime\prime}\times7^{\prime\prime}$
aperture) H$\alpha$ or global [OII] 3727 measurements for global
H$\alpha$ weakens the correlation, but it still reaches $>$3$\sigma$
significance in the B and U bands.

Restoring spiral galaxies brighter than M$_{\rm R}^i=-22.5$ to the sample,
we can fit an EW(H$\alpha$)-correction formula analogous to the
color-correction formula in equation~\ref{eq:colcorr}:
\begin{equation}
\Delta \rm M_{\rm R}^i = 0.5 + 0.026EW(H\alpha).
\end{equation}
For the unweighted inverse fits in Table~\spirtfr, this formula and its B and
U analogues reduce scatter to 0.66, 0.67, and 0.78 mag in the R, B,
and U bands respectively.  The EW(H$\alpha$) correction thus performs
nearly as well as the color correction in both B and R, although it
underperforms in U.

To our knowledge, this paper is the first to report a correlation
between TF residuals and EW(H$\alpha$).
\citet{barton.geller.ea:tully-fisher} search unsuccessfully for such a
correlation using nuclear (fixed slit) H$\alpha$ measurements and
R-band TF residuals.  These results are consistent, given that (a) our
sample yields only a 2.9$\sigma$ correlation between R-band TF
residuals and {\em nuclear} EW(H$\alpha$), and (b) the Barton et al.\
survey targets a specific population (galaxies in close pairs) rather
than a broad cross-section of galaxies.

\subsubsection{Photometric Asymmetry}
\label{sc:photomasym}

The existence of a correlation between TF residuals and photometric
asymmetry has not been firmly established.  \citet{zaritsky.rix:lopsided}
argue that I-band asymmetry and B-band TF residuals are correlated, but
\citet{barton.geller.ea:tully-fisher} suggest that the bulk of this
correlation arises from a systematic error in TF calibration.  We measure
no significant correlation of R-band asymmetry with TF residuals in any
passband, but we do see a marginal 3$\sigma$ correlation between B-band
asymmetry and R-band TF residuals, which weakens for B and U-band
residuals.  We conclude that photometric asymmetry is a weak third
parameter at best.  Since asymmetry arises from multiple causes that may
move TF residuals in different directions (knottiness from dust, unevenness
from recent starbursts or accretion, etc.), sample selection may partly
determine whether or not a correlation is detected.

\subsubsection{Rotation Curve Asymmetry}
\label{sc:rcasym}

Rotation curve asymmetry (defined in Appendix~\ref{sc:asymmeast})
correlates with TF residuals in the B and U bands with $\sim$3$\sigma$
significance.  The RC asymmetry vs.\ TF residual plot
(Figure~\ref{fg:restspircorr}) may also indicate a discrete effect:
galaxies with asymmetry less than $\sim$2\% show lower scatter and a slight
offset relative to those with higher
asymmetry. \citet{barton.geller.ea:tully-fisher} also report a discrete
effect in their close pairs sample: the R-band TF residuals of galaxies
identified as severely kinematically distorted differ from the residuals
for the remaining galaxies by 3$\sigma$ in a K-S test.  We return to the
possible link between kinematic distortions and TF offsets in
\S\ref{sc:voffs}.

\subsubsection{Surface Brightness}
\label{sc:surfbrt}

\citet{courteau.rix:maximal} argue against any surface brightness
dependence in the TF residuals of high surface brightness spiral galaxies,
while \citet{oneil.bothun.ea:red} demonstrate that extreme low surface
brightness (LSB) galaxies can lie well off the TFR.  In our sample we find
no clear correlation in any passband with the most common measure of
surface brightness used in the literature, the extrapolated disk central
surface brightness (as measured from exponential profile fits, R. Jansen,
private communication).  On the other hand, we do detect marginal
correlations between TF residuals and two other measures of surface
brightness.  The observed R-band central surface brightness (measured from
raw profiles including both disk and bulge light, and subject to
uncertainties due to both seeing and the distance to the galaxy) correlates
with U and B-band TF residuals at $>$3$\sigma$, but the detection has high
uncertainty due to a strong central surface brightness--luminosity
correlation (which causes the third parameter correlation to weaken if
weighted TF fits are used, see \S\ref{sc:thirdpartech}), and it disappears
for the full sample.  A more robust quantity, the effective surface
brightness in the U band ($\mu_e^U$, the surface brightness measured {\em
at} $r_e$) displays a 3$\sigma$ correlation with R-band TF residuals, but
this correlation fades in the B and U bands (Figure~\ref{fg:restspircorr}).
However, the correlation becomes stronger for the full sample; we discuss
it in \S\ref{sc:sbeu}.

\subsubsection{Neutral Gas Properties}
\label{sc:neutral}

The present spiral sample shows no correlation between TF residuals
and neutral gas properties such as $M_{\rm HI}$, gas consumption
timescale\footnote{Gas consumption timescales are calculated by
dividing the H\,{\small I} gas mass by the star formation rate, where
gas masses are derived from the catalog H\,{\small I} fluxes of
\citet{bottinelli.gouguenheim.ea:extragalactic} and
\citet{theureau.bottinelli.ea:kinematics}, and star formation rates
are computed from H$\alpha$ fluxes (R.\ Jansen, private communication)
using the calibration of \citet{kennicutt:star} with no correction for
recycling.}, and $M_{\rm HI}/L$, where we determine $L$ in the same
passband as the TF residuals to eliminate any correlation due to
color.  However, we compute gas masses from the catalogs of
\citet{bottinelli.gouguenheim.ea:extragalactic} and
\citet{theureau.bottinelli.ea:kinematics}, and these catalogs do not
provide H\,{\small I} masses for $\sim$31\% of the Sa--Sd subsample.
The incompleteness preferentially affects Sa and peculiar galaxies,
which have the largest TF offsets, so our results should be treated
with caution.

Nonetheless, our failure to detect an $M_{\rm HI}$--TF residual
correlation is interesting in light of the strength of the $M_{\rm
HI}$--luminosity correlation.  For the Sa--Sd subsample, the $M_{\rm
HI}$--luminosity correlation rivals the TFR itself in correlation
strength: although weaker at B and R, it is actually stronger at U,
with a Spearman rank probability of no correlation of $10^{-11}$.
Therefore even a small TF slope error could easily generate a false
correlation between TF residuals and H\,{\small I} mass in this sample.  The fact
that we detect nothing validates our fitting and analysis procedures,
as H\,{\small I} masses and TF velocity widths are measured completely
independently.  Furthermore, since the color--luminosity correlation
is much weaker than the H\,{\small I} mass--luminosity correlation in our data,
this argument supports our claim that slope errors do not drive the
strong correlation we observe between color and TF residuals.

\subsection{What Causes Offsets from the Spiral TFR?}
\label{sc:causes}

The existence of a well-defined Tully-Fisher relation for spiral galaxies
suggests that for large disk galaxies, the total mass of the system (traced
by its rotation velocity) closely couples to the mass of its stars (traced
by their luminosity).  TF residuals may arise from any of three basic
relations underlying the overall relation: (1) the relation between
observed rotation velocity and total mass, (2) the relation between
total mass and stellar mass, and (3) the relation between stellar
mass and observed luminosity.  Below, we examine the possibility that
variations in stellar $M/L$ due to differences in stellar populations can
fully explain the color--TF residual correlation, with TF offsets
interpreted as luminosity offsets.  We then discuss alternative
explanations for the color--TF residual correlation that interpret TF
offsets as velocity width offsets that correlate with color.

\subsubsection{Luminosity Offsets}
\label{sc:loffs}

Variations in stellar $M/L$ due to differences in stellar populations
provide the simplest explanation of TF scatter
\citep[e.g.,][]{rhee:physical}.  Scatter driven by population differences
should be higher in bluer passbands, where variations in $M/L$ are greater,
and indeed TF scatter increases from R to B to U.  If there is negligible
scatter in the correlation between rotation velocity and stellar mass
(i.e., the combination of relations (1) and (2) above), then the
color--TF residual correlation follows directly from the dependence of
color on $M/L$, provided that color variations are not dominated by the
systematic correlation between color and magnitude (in agreement with our
observations, see \S\ref{sc:color} and Figure~\ref{fg:kandcvslum}).  The
only caveat is that different star formation histories may yield different
ratios of luminosity evolution to color evolution ($LE/CE$), producing
different TF offsets for a given color change.  Thus we
can test the hypothesis that stellar population differences drive the
color--TF residual correlation by comparing the slope of the correlation to
the $LE/CE$ predictions of population synthesis models designed to describe
the general population of spiral galaxies.

We perform this comparison using the model predictions of
\citet{bell.:stellar}.  These authors use population synthesis models to
evaluate the slope of the color vs.\ stellar $M/L$ correlation for
passively evolving disk galaxies subject to a variety of star formation
histories in which stars form in a reasonably smooth fashion with no major
bursts.  These slope predictions may be compared to the measured slopes
shown in the top two panels of Figure~\ref{fg:restspircorr}.  The top right
panel shows the color {\em residual}--TF residual correlation, where color
residuals are defined relative to the color--magnitude relation; this
relation offers the most direct measurement of $LE/CE$, suitable for
comparison with a simple model such as a closed box model.  (One might
prefer to define color residuals relative to the color-velocity width
relation; however, because our sample is defined by strict limits in
magnitude rather than in velocity width, the color-velocity width relation
would be biased.)  The standard color--TF residual correlation, uncorrected
for the color--magnitude relation, may correspond better with models
employing mass-dependent star formation histories.

Depending on the particular star formation history used, Bell \& de
Jong's predicted slope varies from 1.8--2.5 for R-band TF residuals
and $B-R$ colors, with a closed box model giving a slope of 2.3.  For
the observed correlation, a forward least-squares fit yields a slope
of 2.2(2.5) $\pm0.5$ for the uncorrected(corrected) correlation, while
a least-squares bisector fit yields 3.5(3.9) $\pm0.5$
\citep[cf.][]{isobe.feigelson.ea:linear}.  The latter fit technique is more
realistic as it assumes intrinsic scatter in both fit variables;
slopes measured by this technique are broadly consistent with the Bell
\& de Jong predictions but appear to favor slightly steeper values.
Slopes measured in the B band are also somewhat steep: e.g. while the
closed box model gives a slope of 3.3 in B, we derive values of 3.2
and 4.4 using forward and least-squares bisector fits, respectively,
for the color residual--TF residual correlation.

The basic agreement between Bell \& de Jong's predictions and our
observations supports our interpretation of the data in terms of stellar
populations.  With this interpretation we find a single
passband-independent stellar-mass TFR (i.e., our ``color-corrected'' TFR,
\S\ref{sc:color}) with lower scatter than the conventional TFR and a form
very similar to that predicted by Bell \& de Jong: $L\propto
V^{4.6\pm0.3}$.  To the extent that there is any discrepancy between
observed and predicted $LE/CE$ ratios, the data would appear to support
slightly steeper slopes than the models.  This result is largely
insensitive to the choice of extinction corrections, even when we omit such
corrections altogether, although it is possible to push the observations
closer to the models with selected analysis techniques (e.g.\ adopting the
type-dependent $q_0$ described in \S\ref{sc:saoffset} and using colors
uncorrected for the color--magnitude relation).  If real, the steeper
slopes might reflect several effects: (1) variations in dark matter
fraction or dark matter structure that correlate with color; (2) systematic
velocity width offsets due to inclination errors or rotation curve
distortions that correlate with color; or (3) star formation histories that
include recent starburst activity ($\la$1 Gyr ago), which would give
$LE/CE$ ratios $\ga$5 (e.g., Bell \& de Jong Figure~5).  In support of
option (3), many NFGS galaxies show morphological and kinematic evidence of
disturbance (\S\ref{sc:pec}), suggesting the likelihood of starbursts
driven by interactions, mergers, or global instabilities.  We discuss
options (1) and (2) in the next section.

We have also analyzed the color--TF residual correlation for one other
complete TF sample, the Ursa Major sample of
\citet{verheijen.sancisi:ursa}, and we find much shallower slopes ($\sim$1
in R, $\sim$2 in B).  These slopes are even shallower than expected purely
from population synthesis.  As the Ursa Major cluster represents a uniform
moderate-density environment, the difference may be related to a greater
homogeneity of star formation histories compared to what we see in the
general field.  In this case the TF residuals in the color--TF residual
correlation might have less to do with luminosity offsets related to
stellar populations and more to do with color-dependent velocity width
offsets driven by one of the mechanisms described in \S\ref{sc:voffs}
below.

\subsubsection{Velocity Width Offsets}
\label{sc:voffs}

TF residuals are very sensitive to small offsets in velocity width because
of the steep slope of the TFR.  Therefore any systematic velocity width
trends in our data must correlate tightly with color, given the success of
the color correction in reducing scatter to near measurement-error levels
in all bands (\S\ref{sc:color}).  Possible sources of velocity width
residuals include variations in dark matter fraction or dark matter
structure, systematic errors in photometric inclination, and symmetric or
asymmetric distortions in rotation curves.  If any of these correlates
strongly with color, it may offer an alternative explanation for the
color--TF residual correlation in terms of velocity width offsets rather
than luminosity offsets.

A full consideration of dark matter structure is beyond the scope of this
paper, but we can address one plausible scenario with straightforward
observable consequences.  Halo contraction in bulge-dominated galaxies
could lead to increased rotation velocities within the central parts of the
galaxies, in turn driving a correlation between color and velocity width
residuals insofar as higher bulge-to-disk ratios correlate with redder
colors.  However, the high color dispersion for later morphological types
\citep{jansen.franx.ea:surface} suggests a relatively weak link between
colors and bulge-to-disk ratios, except for early types such as Sa
galaxies.  Furthermore, the observed RCs for our Sa galaxies do not
decrease in velocity in their outer parts, so we have no independent
evidence that these galaxies' high central concentrations have in fact
increased their measured rotation velocities.  Extended H\,{\small I} data
might reveal such an effect.

Alternatively, velocity width offsets might be driven by photometric
inclination errors that correlate with color.  A color--inclination error
correlation could arise indirectly, via a color-morphology connection
involving either bulge-to-disk ratios or morphological peculiarities, both
of which can affect photometric inclination estimates.  We cannot directly
test this possibility without kinematic inclinations.  However, raising the
inclination cut from $i>40$ to $i>60$ to reduce inclination-related offsets
simply tightens the color--TF residual correlation, reducing its scatter
without significantly changing its slope (in the R band, the range of
acceptable slopes narrows to 2.5--3.4).  Furthermore, we have argued that
inclination errors cannot explain the Sa galaxy offset, although they may
contribute to it (\S\ref{sc:saoffset}), and that the link between colors
and bulge-to-disk ratios is probably fairly weak.  It is also hard to
imagine how a heterogeneous population of peculiar galaxies could yield
consistently high inclinations, as required to explain their high $L$/low
$W^i$ offsets.\footnote{Inclination errors probably do explain a few
individual outliers: for example, tidal elongation may have led to an
incorrect inclination for NGC~5993, the extreme outlier at $-$21.5 in the
color-corrected U-band TFR, Figure~\ref{fg:scattred} (also shown in the top
panel of Figure~\ref{fg:pecgallery}).  Its nominal photometric inclination
is 42 degrees, but its inner parts appear closer to face on.  If its true
inclination were 20 degrees it would shift by $\sim$2.8 mag, bringing it
much more in line with the TFR.  However, NGC~5993 is not at all typical:
its offset does {\em not} follow the color--TF residual correlation.  In
fact, on average, spiral galaxies with peculiarities have slightly lower
apparent inclination than the rest of the spiral sample (63$\degr$ vs.\
66$\degr$).}  Systematically high random errors for peculiar galaxies would
produce a net offset in the wrong direction, due to the nonlinearity of the
sine function.

Rotation curve asymmetries cannot explain the color--TF residual
correlation either, although these asymmetries do correlate weakly with
both TF residuals and $(B-R)_e$ (see \S\ref{sc:rcasym}).  Even supposing that a
5\% asymmetry produces a 5\% offset in rotation velocity (certainly an
overestimate), that velocity offset will in turn generate a TF residual of
only $\sim$0.2 mag. Relatively few galaxies have asymmetries that large
(see Figures~\ref{fg:mfbprune} and~\ref{fg:restspircorr}).

Symmetric rotation curve distortions could be more important, if these
distortions correlate with color.  2D velocity fields would be required to
evaluate their role.  \citet{franx.:elongated} have shown that
intrinsically elongated disks can produce symmetric kinematic distortions,
but random orientations to the line of sight will lead primarily to scatter
rather than to a systematic correlation.  Galaxy interactions could create
distortions that correlate with blue color, but a mechanism for shifting
both radio and optical velocity widths in the same direction is not
obvious.  \citet{tutui.sofue:effects} argue that tidal interactions broaden
H\,{\small I} linewidths, which would actually weaken the color--TF
residual correlation for the H\,{\small I} TFR, as would H\,{\small I}
confusion.  Technically, these authors' observations do not distinguish
between broadening of H\,{\small I} linewidths and narrowing of CO
linewidths, which is likely to occur as molecular gas concentrates at the
center of a galaxy where it does not sample the full velocity field
\citep{mihos.hernquist:gasdynamics}.  Ionized gas is also likely to
concentrate at the center for starburst galaxies, yielding low velocity
width TF outliers due to rotation curve truncation \citep[][see also
\S\ref{sc:allmorphthirdpars}]{barton.geller.ea:tully-fisher,barton.:possible}.
While this effect may create offsets for a few galaxies, it does not
contribute to systematic trends: TF residuals show no significant
correlation with rotation curve extent.

We conclude that the color--TF residual correlation is most easily
explained as being fundamentally a color--luminosity residual correlation,
although it may be steepened by a secondary color--velocity width residual
correlation.

\section{The TFR for the General Galaxy Population}
\label{sc:fulltfr}

\begin{figure}[tb]
\epsscale{0.5}
\plotone{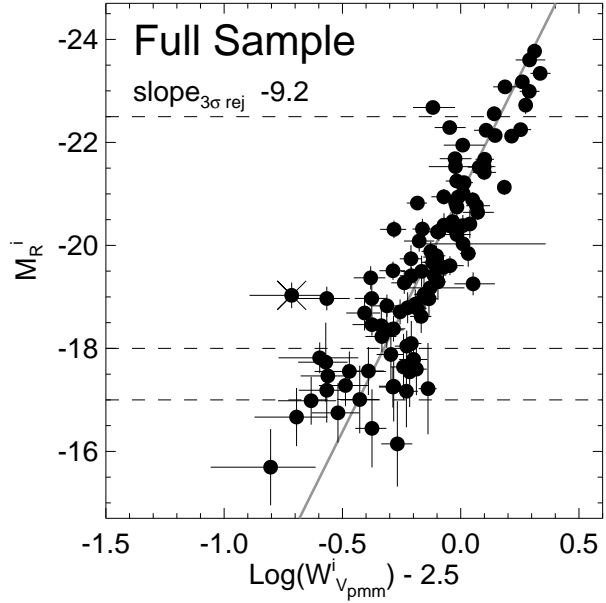}
\caption{TFR for the full sample, including all morphologies and
luminosities but restricted to $i>40$.  Dashed lines indicate
luminosity cuts used for Spearman rank tests
(\S\ref{sc:allmorphthirdpars}--\ref{sc:dwarfthirdpars}, see also
\S\ref{sc:thirdpartech}).  The gray line shows an unweighted inverse
fit to all points, and the X marks a point automatically rejected by
that fit as a $>$3$\sigma$ outlier.}
\label{fg:allgalsone}
\end{figure}

We now turn to the full TF sample of NFGS emission-line galaxies,
with no restrictions on luminosity or morphology, but still requiring
$i>40$.  Table~\fulltfr\ gives fitted TF parameters for this sample in
the U, B, and R bands, with optical RC results based on 108 galaxies
(107 at U) and H\,{\small I} linewidth results based on 76 galaxies (75 at U).
The R-band TFR is shown in Figure~\ref{fg:allgalsone}.

\begin{deluxetable}{rrrrr}
\tablenum{\fulltfr}
\tabletypesize{\footnotesize}
\tablewidth{0pt}
\tablecaption{Tully-Fisher Fits to the Full Sample~\tablenotemark{a}}
\tablehead{ & & RC Results & & H\,{\scriptsize I} Results \\
\cline{2-4} \cline{5-5}
\\
\colhead{Band} &
\colhead{Wtd Inv} &
\colhead{Bivariate} &
\colhead{Unwtd Inv} &
\colhead{Unwtd Inv} }
\startdata
\cutinhead{Slope} \\
U &  -9.47$\pm$0.21 &  -7.91$\pm$0.24 &  -8.97$\pm$0.22 &  -8.48$\pm$0.16 \\
B &  -9.65$\pm$0.21 &  -8.31$\pm$0.25 &  -9.06$\pm$0.21 &  -8.48$\pm$0.16 \\
R &  -9.92$\pm$0.21 &  -8.77$\pm$0.24 &  -9.23$\pm$0.22 &  -8.56$\pm$0.16 \\
\cutinhead{Zero Point} \\
U & -19.68$\pm$0.04 & -19.79$\pm$0.04 & -20.00$\pm$0.04 & -19.87$\pm$0.03 \\
B & -19.76$\pm$0.04 & -19.86$\pm$0.04 & -20.02$\pm$0.04 & -19.84$\pm$0.03 \\
R & -20.72$\pm$0.04 & -20.82$\pm$0.04 & -20.99$\pm$0.04 & -20.80$\pm$0.03 \\
\cutinhead{Scatter~\tablenotemark{b}} \\
U & 1.26(0.78) & 1.05(0.68) & 1.19(0.75) & 1.09(0.53) \\
B & 1.20(0.79) & 1.02(0.70) & 1.11(0.75) & 1.03(0.53) \\
R & 1.16(0.81) & 1.02(0.73) & 1.07(0.76) & 1.00(0.53) \\
\enddata
\tablenotetext{a}{Fit results from weighted inverse, bivariate, and
unweighted inverse fitting techniques for optical RC (W$_{V_{pmm}}$) and
radio (W$_{\rm HI}$) T-F calibrations (see Appendix~\ref{sc:fitting}).  The
functional form of the TFR is M$_{\lambda}^i={ zero\, point}+
slope(\log{({\rm W}^i)} - 2.5)$.  Errors given are the formal statistical
errors from a two-step fit, see Appendix~\ref{sc:fitting}.  We require $i>40$.
The samples used for the different optical RC fits vary slightly, in that
one galaxy has no U-band data.}
\tablenotetext{b}{Measured biweight scatter and predicted scatter
(in parentheses) from measurement errors.}
\end{deluxetable}

\begin{figure}[tb]
\epsscale{0.8}
\plotone{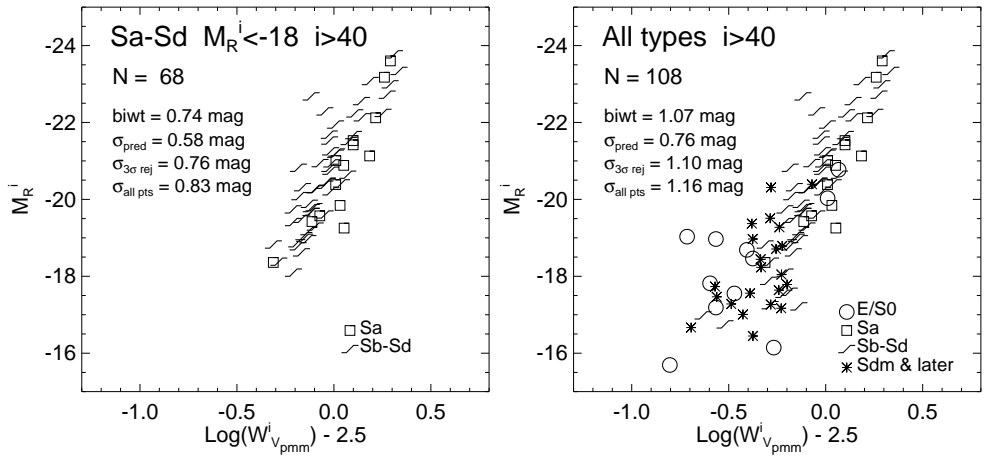}
\caption{Increase in TF scatter caused by extending the sample in
morphology and luminosity. Scatter measurements are relative to an
unweighted inverse fit.  The full sample includes two emission-line
galaxies of type E, whose inclinations may be unreliable.  We include them
for completeness, noting that one is very disky.}
\label{fg:addback}
\end{figure}

As illustrated in Figure~\ref{fg:addback}, scatter is dramatically higher
for the full sample than for the spiral sample of
\S\ref{sc:spiralcal}--\ref{sc:spiral3}.  Including non-spiral and dwarf
galaxies adds 0.77 mag of scatter in quadrature, yielding a total scatter
of 1.07 mag.  Faint non-spiral galaxies are responsible for most of the new
scatter, so the relative importance of the morphology and luminosity
extensions is unclear.  The two E/S0's brighter than M$^i_{\rm R}=-20$
behave much like Sa galaxies, with a slight low $L$/high $W^i$ offset.
\citet{neistein.maoz.ea:tully-fisher} observe a similar TF offset for S0
galaxies using stellar kinematics.

\clearpage

\subsection{Physical Sources of Scatter in a Morphology-Blind TFR}
\label{sc:allmorphthirdpars}

We begin by broadening the spiral sample of \S\ref{sc:spiral3} to
include all morphologies, staying within the luminosity limits used in
\S\ref{sc:spiral3} (Table~\lumcuts\ and Figure~\ref{fg:allgalsone})
and as usual requiring $i>40$.  In \S\ref{sc:dwarfthirdpars} we extend
the luminosity range 1~mag fainter.  Adding non-spirals brighter than
M$_{\rm R}^i=-18$ to the spiral sample strengthens the correlations
between TF residuals and both rotation curve asymmetry and nuclear
EW(H$\alpha$).  We also see a hint of a new correlation between TF
residuals and gas consumption timescale (based on incomplete data, as
discussed in \S\ref{sc:neutral}, but with 88\% completeness for the
newly added non-spiral galaxies).  The TF residual--$\mu_e^U$
correlation remains about the same.

\begin{figure}[tb]
\epsscale{0.35}
\plotone{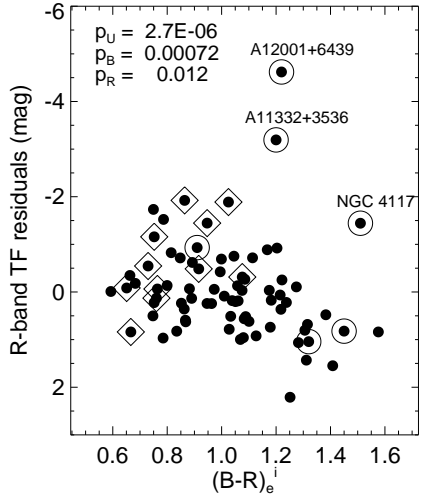}
\caption{Correlation between TF residuals and $(B-R)_e^i$ color for
galaxies of all morphologies in the range $-18<\rm M^i_{\rm R}<-22.5$.
As before, we consider only galaxies with $i>40$.  Boxed and circled
points represent galaxies with morphological type later than Sd and
earlier than Sa, respectively, which were not included in the spiral
sample of \S\ref{sc:spiral3}.  Most of these galaxies follow the
spiral correlation well, except for three clear S0 outliers (discussed
in \S\ref{sc:allmorphthirdpars}).  The Spearman rank probability of no
correlation between $(B-R)_e^i$ and TF residuals in each optical band
is given as $p_{band}$.  Only R-band TF residuals are plotted.}
\label{fg:allcorrstdlumcuts}
\end{figure}

However, the expanded sample displays {\em weaker} correlations between TF
residuals and both effective $B-R$ color and global (as opposed to nuclear)
EW(H$\alpha$).  As shown in Figure~\ref{fg:allcorrstdlumcuts}, the
$(B-R)_e$ correlation actually falls below 3$\sigma$ at R.  Although most
non-spiral galaxies, marked with circled and boxed points in
Figure~\ref{fg:allcorrstdlumcuts}, follow the correlation established by
spiral galaxies quite well (albeit with slightly larger scatter), three of
the six emission-line E/S0 galaxies in the sample are extreme outliers.
These three outliers, labeled by name in Figure~\ref{fg:allcorrstdlumcuts},
have red colors $\ga$1.2 and TF residuals $>$1 mag toward high $L$/low
$W^i$.  H\,{\small I} linewidths are available for two of the three S0
outliers; these two galaxies are clear outliers from the H\,{\small I} TFR
as well, although the H\,{\small I} residuals are smaller than the optical
RC residuals.

The large TF residuals for these three S0 galaxies suggest that our
observations do not accurately reflect their disk rotation velocities
and/or their underlying luminosities.  In each case the data are likely to
be influenced by central activity, in the form of either an AGN or a
central starburst.  However, the central luminosity enhancements are too
small to account for the TF offsets. Inclination errors are also unlikely
to explain the outliers, as all three galaxies show disky morphologies and
inclinations of $\sim$60-70 degrees.  We suspect kinematic effects.
NGC~4117 is a Seyfert II and has a truncated rotation curve that extends to
only 0.56$r_e$; its stellar kinematics are peculiar and may indicate
gas-stellar decoupling.  A11332$+$3536 hosts a central starburst and has a
truncated rotation curve that extends to 0.92$r_e$; the gas seems to lie in
an inclined counterrotating disk
\citep{kannappan.fabricant:broad}. A12001$+$6439 has an unusually bright
nucleus and its rotation curve reaches only 1.31$r_e$; the gas displays
velocity reversals.  For the latter two galaxies H\,{\small I} masses are
available, from which we calculate rapid gas consumption timescales of
$\sim$2 Gyr.

The irregular kinematics and starburst/AGN activity found in these S0
galaxies may result from recent interactions or mergers.  As discussed by
\citet{barton.:possible}, interacting galaxies may display radially
truncated emission due to centrally concentrated starbursts, leading to
artificially low velocity widths.  Interactions and mergers can also
generate non-coplanar gas and stellar rotation in early-type galaxies
\citep{kannappan.fabricant:broad,haynes.jore.ea:kinematic}, in which case
artificially low gas velocities will be measured because the spectrograph
slit P.A. is aligned to the stellar major axis rather than to the gas major
axis.  Deviations from the color--TF residual correlation provide a
convenient and physically motivated way to identify such galaxies and prune
them from a TF sample.

\subsection{Physical Sources of Scatter in the Dwarf Extension of the TFR}
\label{sc:dwarfthirdpars}

\begin{figure}[tb]
\epsscale{0.45}
\plotone{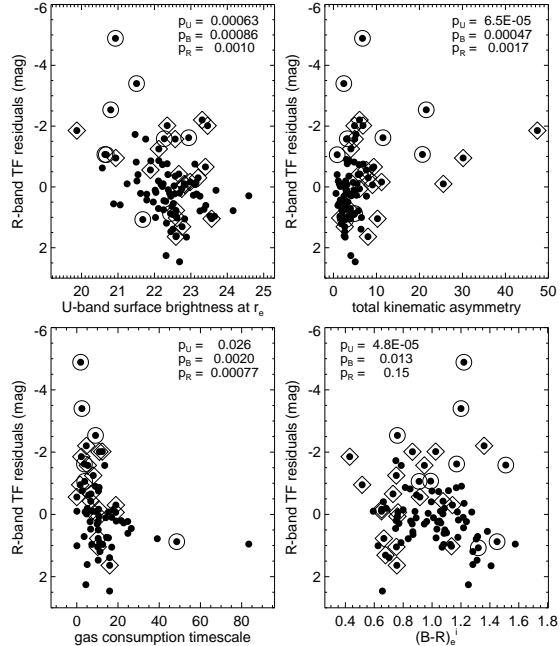}
\caption{TF residuals vs.\ physical properties of the full sample (all
morphologies, $i>40$, $-17<\rm M^i_{\rm R}<-22.5$, see
\S\ref{sc:dwarfthirdpars} and Figure~\ref{fg:allgalsone}).  Boxed and
circled points represent galaxies with morphological type later than
Sd and earlier than Sa, respectively, which were not included in the
spiral sample of \S\ref{sc:spiral3}.  (Other additional points are
dwarf late-type spiral galaxies.)  The Spearman rank probability of no
correlation between the parameter and TF residuals in each optical
band is given as $p_{band}$.  Only R-band TF residuals are plotted.}
\label{fg:fullsampcorr}
\end{figure}

Figure~\ref{fg:fullsampcorr} plots third parameter--TF residual
correlations for the full TF sample, with luminosity cuts extended to
include dwarf galaxies down to M$_{\rm R}^i=-17$ (see Table~\lumcuts).
Adding these dwarfs dilutes the third-parameter correlations for
$(B-R)_e$ and EW(H$\alpha$) severely; for R-band TF residuals they
become scatterplots.  The only third-parameter correlations that reach
3$\sigma$ significance in the R band involve parameters that showed
weak or nonexistent correlations for the spiral sample: rotation curve
asymmetry, U-band effective surface brightness, and gas consumption
timescale (see \S\ref{sc:rcasym}--\ref{sc:neutral}).

\begin{figure}[tb]
\epsscale{0.8}
\plotone{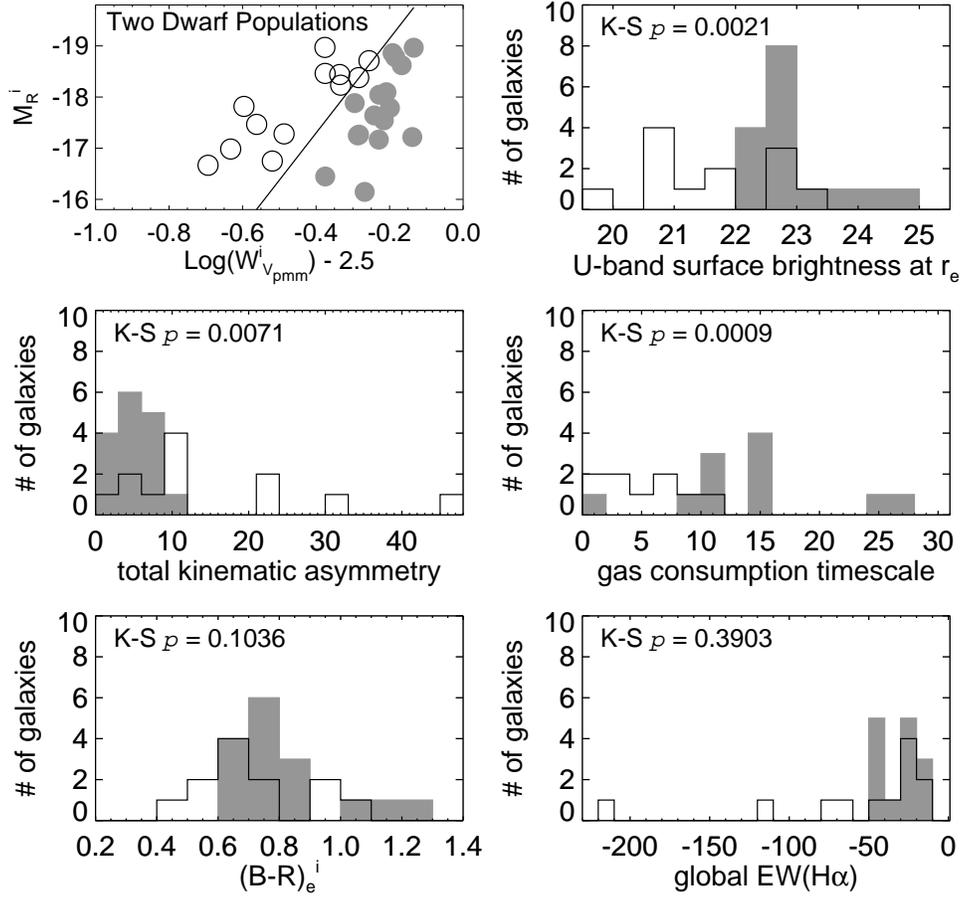}
\caption{Physical properties of dwarf galaxies on either side of the
TFR.  Dwarfs on the high $W^i$/low $L$ side of the TFR are shown in
gray, while those on the low $W^i$/high $L$ side are outlined in
black.  Each panel lists the Kolmogorov-Smirnov probability that the
two sets of dwarfs were drawn from the same parent population with
respect to the property shown.  Here ``dwarf'' means anything fainter
than M$_{\rm R}^i=-19$; the luminosity cuts used for the
third-parameter tests are {\em not} applied.  Galaxies with truncated
RCs ($<$1.3$r_e$) have been excluded from the figure because of
uncertain location with respect to the TFR (however they do appear in
Figure~\ref{fg:fullsampcorr}).  Restoring these galaxies minimally
affects the K-S test results for gas consumption timescale and U-band
surface brightness but weakens the results for rotation curve
asymmetry to 2$\sigma$ and eliminates any significant result for $B-R$
color.  The reference TFR is the full sample TFR shown in
Figure~\ref{fg:allgalsone}.}
\label{fg:twodwarfs}
\end{figure}

Because these correlations are highly scattered, they are more easily
understood in terms of bulk differences in properties for galaxies on
either side of the TFR, rather than in terms of a linear trend like the
color--TF residual correlation.  Figure~\ref{fg:twodwarfs} illustrates these
bulk differences with separate histograms for galaxies on either side of
the TFR, considering only galaxies fainter than M$_{\rm R}^i=-19$ since
these galaxies are the primary drivers of the new correlations.  (Unlike
Figure~\ref{fg:fullsampcorr}, Figure~\ref{fg:twodwarfs} excludes galaxies
with truncated rotation curves, which might appear on the low $W^i$ side of
the TFR erroneously; however, Figure~\ref{fg:twodwarfs} applies no lower
limit on luminosity.)  In what follows we refer to galaxies fainter than
M$_{\rm R}^i=-19$ as ``dwarfs.''  Figure~\ref{fg:addback} shows that the
non-spiral morphologies typical of dwarfs are most common below M$_{\rm
R}^i=-19$.

\subsubsection{Rotation Curve Asymmetry}
\label{sc:dwarfrcasym}

Asymmetries in rotation curves arise from multiple sources including
turbulence, bars, tidal distortions, early or late-stage infall, and
satellite accretion.  Most of these underlying phenomena are likely to
enhance star formation, qualitatively explaining the sense of the
correlation in Figure~\ref{fg:fullsampcorr} as well as its stronger
statistical signal in bluer passbands.  However, the scatter and
substructure in the correlation suggest that the RC asymmetry--star
formation connection may be complex.  On the the high $L$/low $W^i$
side of the TFR, several dwarf galaxies show RC asymmetries $\ga$10\%
(much higher than expected from the RC asymmetry--luminosity relation
in Figure~\ref{fg:kandcvslum}). On the other side of the TFR, RC
asymmetries are generally lower and more uniform
(Figure~\ref{fg:twodwarfs}; see also \S\ref{sc:dwarfreal}).

\clearpage

\subsubsection{U-Band Effective Surface Brightness}
\label{sc:sbeu}

If U-band effective surface brightness ($\mu_e^U$) measures the
surface density of recent star formation in galaxy disks, then high
$\mu_e^U$ (numerically small values) should correlate with high global
star formation activity, consistent with the observed TF
residual--$\mu_e^U$ correlation.  However this view may be too
simplistic, because galaxies ranging from undisturbed spirals to
emission-line S0's with kinematically decoupled gas all seem to follow
the same correlation, albeit with very large scatter.  Given the
constancy of the Spearman test significance for this correlation
whether using U, B, or R-band TF residuals
(Figure~\ref{fg:fullsampcorr}), we suggest that variations in $\mu_e^U$
may track not only star formation but also structural differences
between classes of galaxies located in different parts of the
$L$--$W^i$ plane, such as differences in the concentration of light or
the velocity width at a given luminosity.  Dwarfs on opposite sides of
the TFR have different $\mu_e^U$ distributions
(Figure~\ref{fg:twodwarfs}), analogous to the different rotation curve
asymmetry distributions discussed in the previous section.  While low
to moderate surface brightnesses occur on both sides of the TFR, high
surface brightnesses occur only on the high $L$/low $W^i$ side.

\subsubsection{Gas Consumption Timescale}
\label{sc:gastime}

Typical gas consumption timescales also appear to differ for dwarfs on the
two sides the TFR, although this result relies on incomplete H\,{\small I}
mass data (82\% complete for the galaxies in Figure~\ref{fg:twodwarfs},
with fainter galaxies preferentially missing).  In
Figures~\ref{fg:fullsampcorr} and~\ref{fg:twodwarfs}, only a few galaxies
have low $L$/high $W^i$ (positive) residuals and very short gas consumption
timescales.  By contrast, short gas consumption timescales are the norm for
galaxies with negative residuals.  Most of the latter set are
low-luminosity non-spiral galaxies, which explains the emergence of a gas
consumption timescale correlation in the full sample, where there was no
correlation in the spiral sample of \S\ref{sc:spiral3}.

\subsection{Is the Dwarf Split Real?}
\label{sc:dwarfreal}

In \S\ref{sc:dwarfrcasym}--\ref{sc:gastime} we showed that if we divide the
dwarf population into two groups with positive and negative offsets from
the TFR, the groups have distinct physical properties (summarized in
Figure~\ref{fg:twodwarfs}).  Although K-S tests yield only $\sim$3$\sigma$
confidence that the two populations are distinct based on any one property,
the same split is observed for three completely independent parameters:
rotation curve asymmetry (based on high-resolution spectra), $\mu_e^U$
(based on surface photometry), and gas consumption timescale (based on
catalog H\,{\small I} data and NFGS spectrophotometry).  The TF residuals
for the two dwarf populations differ by $\sim$2 mag in the mean.

We now consider whether systematic effects might create such a split.
Because the two sets of dwarfs in Figure~\ref{fg:twodwarfs} show similar
distributions in inclination and morphology, correlated inclination errors
are unlikely.  Truncated optical RCs are also unlikely to play a role, as
Figure~\ref{fg:twodwarfs} excludes galaxies whose RCs do not reach
1.3$r_e$.  Furthermore, replacing optical velocity widths with H\,{\small
I} linewidths preserves the division.  RC asymmetries may affect measured
velocity widths, but the asymmetries we measure are too small to produce a
2 mag difference between the two groups of dwarfs (cf.\ \S\ref{sc:voffs}).

Correlated distance errors might be a concern.  Both dwarf
populations show broad distributions in R.A.-Dec.-redshift space, but due
to the NFGS selection procedure, the two sets of dwarfs in
Figure~\ref{fg:twodwarfs} have systematically different recession
velocities: 500--1300 and 700--1600 $\rm km \, s^{-1}$ for underluminous
and overluminous dwarfs respectively, with over half the underluminous set
in the range 500--700 $\rm km \, s^{-1}$ (measured with respect to the
Local Group).  To test whether peculiar velocities or local flows may have
affected our distance estimates and produced a spurious bifurcation in
luminosities, we have substituted distances estimated by a number of
techniques.  These include: (1) Hubble-law distances assigned using group
recession velocities rather than individual galaxy recession velocities
\citep[][]{tully:nearby,
tully.shaya.ea:nearby};\footnote{\label{fn:umoutlier} This substitution
yields one large shift (0.6 mag toward higher luminosity), for a galaxy
(UGC~6446) assigned ``questionable'' membership in the Ursa Major Cluster
by \citet{tully.verheijen.ea:ursa}.  Its projected location lies near the
cluster outskirts and its velocity approaches the cluster lower limit.} (2)
absolute distances derived from the light-to-mass model of
\citet{tully.shaya.ea:nearby}; (3) Hubble-law distances with and without
corrections for Virgocentric infall (\S\ref{sc:absmag}) and/or for the
Local Anomaly \citep{han.mould:velocity}.  The segregation of the two dwarf
populations is robust under all of these substitutions, with the caveat
that many of the overluminous (low velocity width) dwarf galaxies lack
distances from methods (1) and (2).

We conclude that the TF offset between the two groups of NFGS dwarfs is
probably real.  These results confirm trends reported for two other
heterogeneous TF samples \citep{stil.israel:faint,pierini:internal}.  Both
studies identify a rotationally supported dwarf population on the faint
side of the TFR and a population with comparable rotational and random
velocities on the bright side (where we measure high rotation curve
asymmetries).

The physical mechanisms responsible for the disturbed kinematics and rapid
gas consumption of the dwarfs on the bright side of the TFR are uncertain,
but interactions are an obvious possibility.  Without deep wide-field
imaging we cannot evaluate the frequency of close companions for the two
populations, particularly given the possibility of relatively faint
neighbors.  Nonetheless, it is intriguing to note that deep survey work by
\citet{taylor.thomas.ea:survey} has demonstrated a statistically higher
rate of companions near starbursting dwarfs than near LSB dwarfs \citep[see
also][]{pustilnik.kniazev.ea:environment}.  The analogy is suggestive
rather than exact, as our two dwarf populations do not show a definitive
split in EW(H$\alpha$), although the three most extreme starbursting dwarfs
do fall on the bright side of the TFR (Figure~\ref{fg:twodwarfs}).

\subsection{Dwarfs in Ursa Major}
\label{sc:umdwarfs}

The three studies that have reported a difference in dwarf properties on
either side of the TFR are all based on heterogeneous galaxy samples
representing a variety of environments \citep[our
own;][]{stil.israel:faint,pierini:internal}.  Here we investigate dwarf
properties and their relationship to the TFR for the Ursa Major sample of
\citet[][]{verheijen.sancisi:ursa}, which represents a single
moderate-density environment.  While the VS data set does not include
kinematic asymmetries or U-band data, we can examine B-band surface
brightnesses and TF offsets.  (Although not significant at the 3$\sigma$
level, $\mu_e^B$ trends basically follow $\mu_e^U$ trends with weaker
signal.)

\begin{figure}[tb]
\epsscale{0.5}
\plotone{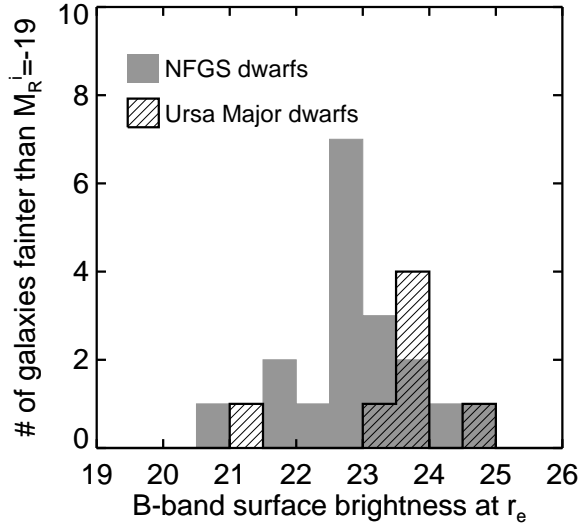}
\caption{Comparison of B-band effective surface brightnesses for the
NFGS and Ursa Major dwarfs that meet the TF sample criteria described
in \S\ref{sc:umdwarfs}.  To match the Ursa Major sample completeness
limit (M$_{\rm B}=-16.8$), the figure omits many of the fainter NFGS
galaxies included in \S\ref{sc:dwarfthirdpars}.  We measure $\mu_e^B$
for the Ursa Major galaxies directly from the profiles given in
\citet{tully.verheijen.ea:ursa}, using the effective radii listed in
their Table~4 and the Dexter java applet in the ADS (NASA's
Astrophysics Data System).  The measured values are corrected for a
small mean offset (0.2 mag arcsec$^{-2}$) determined by comparing
$\mu_e^B$ values for six galaxies common to both the NFGS and Ursa
Major databases; without this correction the dwarf surface brightness
offset shown in the figure would be smaller.}
\label{fg:sbcomparison}
\end{figure}

As noted in \S\ref{sc:cluster} (see also Figure~\ref{fg:verhfigscatt}), the
Ursa Major TFR appears to diverge from the NFGS TFR at the faint end.  When
we expand the two spiral samples to include late-type galaxies and dwarfs
down to M$_{\rm B}=-16.8$ (the Ursa Major sample completeness limit), we
find a TF offset of $\sim$0.5 mag between the NFGS and Ursa Major dwarfs,
with the Ursa Major dwarfs showing higher $L$/lower $W^i$.\footnote{Three
of VS's Sdm and Sm galaxies have obviously discrepant W$_{50}$ values
(NGC~3985, UGC~7089, and NGC~4218), so for these we substitute W$_{20}$ $-$
20 $\rm km \,s^{-1}$ (cf.\ \S\ref{sc:vcalcs}).}  For the same NFGS and Ursa
Major dwarfs, we also find a mean effective surface brightness ($\mu_e^B$)
offset of $\sim$0.5 mag arcsec$^{-2}$, with the Ursa Major dwarfs {\em
fainter}.  These results do not conform to what we observe in the NFGS,
where overluminous dwarfs have relatively bright $\mu_e^B$ values.  Also,
Figure~\ref{fg:sbcomparison} shows that the Ursa Major dwarfs display a
very narrow range in $\mu_e^B$ compared to the NFGS, with a faint median
value of 23.7.  Herein lies a possible explanation for the discrepancy: if
LSB dwarfs are more likely to have rising RCs at their outermost observed
radii, then the ``true'' halo velocities of the Ursa Major dwarfs may be
systematically underestimated \citep[cf.][]{verheijen:ursa*1}.

In this case the divergence of the two TFRs may be due to real physical
differences between the two dwarf populations, and the narrow range of
properties seen for the Ursa Major dwarfs may simply reflect the specific
environment of Ursa Major.  The cluster consists of a moderate overdensity
of spiral galaxies that lacks any discernible core, contains no elliptical
galaxies, and includes surprisingly few dwarfs relative to bright galaxies
\citep{tully.verheijen.ea:ursa, trentham.tully.ea:ursa}.
\citet{trentham.tully.ea:ursa} infer from the scarcity of dwarf galaxies in
Ursa Major that the cluster collapsed late, so that its dwarf halos formed
after reionization when conditions were poor for star formation.

Nonetheless, the environmental differences hypothesis should be
treated with caution, because the Ursa Major dwarfs differ from those
in the NFGS at only the 2$\sigma$ level, with velocity width data
missing for a few dwarfs and with the cluster membership of another
dwarf in question (see note~\ref{fn:umoutlier}; this galaxy happens to
appear in both data sets).  Furthermore, we cannot completely
eliminate small systematic uncertainties in relative distance scale
and photometric calibration between the two data sets; a larger and
more uniform data set would be required to definitively establish the
existence of environmental effects.

\subsection{Two Dwarf Galaxy Populations and the Slope of the TFR}
\label{sc:dwarfslope}

\begin{figure}[tb]
\epsscale{0.75}
\plotone{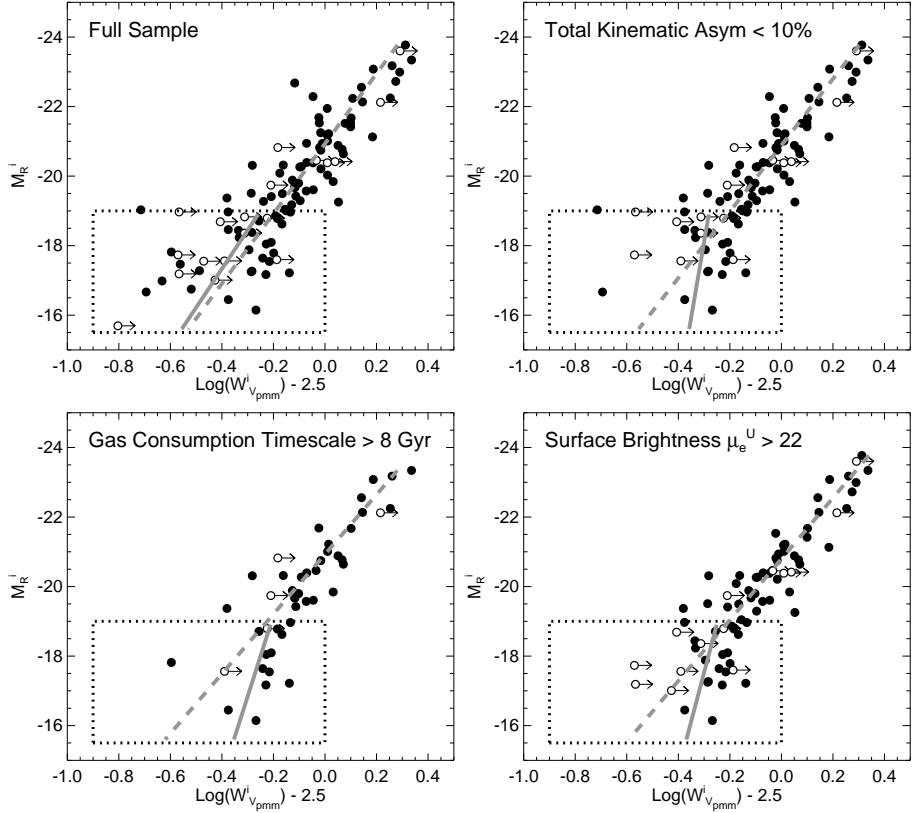}
\caption{Effects of sample selection on the faint-end slope of the
TFR.  Each panel shows galaxies that meet the selection criterion
indicated and have inclinations $i>40$.  Arrows mark galaxies whose
rotation curves are truncated at less than 1.3$r_e$; these galaxies
may have velocity widths larger than we measure.  The dashed and solid
gray lines show the TFR for galaxies brighter and fainter than M$_{\rm
R}^i=-19$, respectively; the dashed line determined by the brighter
galaxies extends into the box for reference only.}
\label{fg:slopebreak}
\end{figure}

As noted by \citet{pierini:internal}, the existence of two dwarf
populations complicates TF slope measurements.  For the full
sample, we find a linear TF slope with no break at low luminosities
(top left panel of Figure~\ref{fg:slopebreak}).  However, different
sample selection criteria may favor one dwarf population over the
other, with consequences for the faint-end slope.

If we consider the TFR as a physical rather than an empirical
relation, linking luminosity and rotation velocity for passively
evolving disk galaxies, then we should restrict it to undisturbed
rotationally supported systems.  Dwarf galaxies on the low $L$/high
$W^i$ side of the TFR best meet this description.  These galaxies show
moderate rotation curve asymmetries, indicating that they are probably
rotationally supported.  In addition, their long gas consumption
timescales and low U-band effective surface brightnesses appear
consistent with quiescent evolution.  (We note that photometric
``normalcy'' is not a useful criterion for dwarfs, as almost all are
lumpy and asymmetric, so defining what is ``peculiar'' becomes
extremely difficult.)

Figure~\ref{fg:slopebreak} demonstrates the effect of selecting restricted
TF samples to match the properties of the quiescent dwarfs at all
luminosities.  Each panel shows separate TF fits for galaxies above and
below M$_{\rm R}^i=-19$, first for the full TF sample, and then for
subsamples restricted in turn to have low kinematic asymmetry, long gas
consumption timescale, and faint U-band effective surface brightness.  The
faint-end fits should not be interpreted as precisely measuring the slope
of the dwarf TFR, given the small number statistics and short lever arm.
Nonetheless, these fits clearly suggest a faint-end slope discontinuity for
TF samples restricted to undisturbed, rotationally supported galaxies: the
dwarfs in these samples appear underluminous when compared to the
extrapolated bright-galaxy TFR.

Breaks in the faint-end slope of the TFR have been previously observed
for extremely late type spirals \citep{matthews..ea:exploration} and
LSB dwarfs \citep{mcgaugh.schombert.ea:baryonic}.  McGaugh et al.\
demonstrate that a straight TF slope (the ``baryonic Tully-Fisher
relation'') can be restored for LSB dwarfs by mathematically
converting their gas into stars and correcting the luminosity for
these unformed stars.  This procedure does not work very well for our
faint galaxies, but McGaugh et al.\ see significant effects primarily
for galaxies fainter than our cutoff of M$_{\rm R}^i\sim-16$.

\section{Conclusions}
\label{sc:concltf1}

The Nearby Field Galaxy Survey offers an ideal sample for exploring
the link between offsets from the Tully-Fisher relation and the
evolutionary states of galaxies.  Galaxies in the NFGS were selected
from the CfA~1 redshift survey without preference for morphology or
environment and span a wide range of luminosities, $-23< \rm M_{\rm
B}<-15$.  Using UBR photometry, optical rotation curve data, and H\,{\small I}
linewidths, we have analyzed TF residuals for two subsamples of the
NFGS, a sample of Sa--Sd galaxies brighter than $\rm M_{\rm R}^i=-18$
and an extended sample including dwarfs and non-spiral galaxies.

Within the spiral sample, we find strong third-parameter correlations
between TF residuals and both $B-R$ color and EW(H$\alpha$). The
cleanest correlations are achieved using effective colors (measured
within the effective radius) and global equivalent widths of H$\alpha$
and [OII] (integrated over the entire galaxy).  Using the color--TF
residual correlation in each TF band to define a color-correction term
to the TFR, we reduce scatter to a single constant value across the R,
B, and U bands, eliminating the usual trend of increasing scatter with
bluer passbands.  The remaining scatter of $\sim$0.65 mag (at a slope
of $-$10) approximately equals the scatter we expect from measurement
errors.  An EW(H$\alpha$)-correction term performs equally well for
the R and B-band TFRs and almost as well for the U-band TFR.

The color and EW(H$\alpha$) correlations are continuous, but their
statistical signal is strongly driven by two morphology classes: peculiar
galaxies (by which we mean recognizable spiral galaxies with oddities such
as warps, multiple nuclei, or interacting companions) that fall on the high
luminosity/low velocity width side of the TFR, and Sa galaxies that fall on
the low luminosity/high velocity width side of the TFR.  This result offers
a different perspective on the longstanding debate over the Sa offset, by
linking the offset simply to the fact that Sa galaxies are consistently
red.

Differences in star formation history offer the simplest explanation for
the color--TF residual correlation: as stellar populations evolve, changes
in their colors and luminosities are intimately connected.  We argue that
the steep slope of the correlation cannot be explained in terms of a
trivial correlation due to passband-dependent scatter.  Interpreting TF
residuals as luminosity residuals, the slope of the color--TF residual
correlation can be directly compared with the predictions of population
synthesis models of disk galaxy evolution.  The slopes we find using R and
B band TF residuals closely match the slopes predicted from the passively
evolving galaxy models of \citet{bell.:stellar}.  However, we find a slight
suggestion of steeper slopes, which might be taken as evidence that in
addition to passive evolution, many spiral galaxies experience significant
starburst-driven evolution, e.g.\ triggered by interactions or minor
mergers (as independently suggested by the disturbed morphology and
kinematics of many of our bluer galaxies).  Variations in dark matter
structure could also steepen the correlation if they were correlated with
color, although we find no independent evidence for such an effect.
Finally, the present data set cannot conclusively rule out systematic
trends caused by kinematic distortions or inclination errors that correlate
with galaxy color: such trends are unlikely to explain the entire color--TF
residual correlation but may affect its slope.

Removing the spiral morphology restriction, we find that most
non-spiral galaxies brighter than $\rm M_{\rm R}^i=-18$ follow the
color--TF residual and EW(H$\alpha$)--TF residual correlations
established by spiral galaxies, but with greater scatter.  A few
emission-line S0 galaxies at the faint end of the included range
deviate strongly from both the TFR and the color--TF residual
correlation; these galaxies all show starburst or AGN activity and
anomalous gas kinematics.

Dwarf galaxies fainter than $\sim\rm M_{\rm R}^i=-18$ do not follow
the color--TF residual correlation.  However, dwarfs on the high
luminosity/low velocity width side of the TFR have higher rotation
curve asymmetries, higher U-band effective surface brightnesses, and
shorter gas consumption timescales than dwarfs on the low
luminosity/high velocity width side of the TFR.  These properties
suggest that many of the high $L$/low $V$ dwarfs are ``disturbed,''
while the low $L$/high $V$ dwarfs are likely to be passively evolving,
rotationally supported systems.

This split in dwarf properties implies that the faint-end slope of the TFR
depends on sample selection.  We find no break in slope for dwarf galaxies
in the full NFGS sample.  However, if we select for rotationally supported,
passively evolving galaxies at all luminosities, we find evidence for a
break toward steeper slope at the faint end of the TFR, consistent with
recent work by \citet{matthews..ea:exploration} and
\citet{mcgaugh.schombert.ea:baryonic}.

\acknowledgements We thank Eric Bell for performing additional model
calculations to answer our questions and for pointing us to helpful
references.  We also thank Betsy Barton for sharing information on
technical aspects of Tully-Fisher analysis, and Rolf Jansen for providing
both data and general expertise.  Finally, we are grateful to Margaret
Geller, Douglas Mar, and Kristin Nelson-Patel for helpful discussions, and
to John Huchra, Renzo Sancisi, and the anonymous referee for critical
readings of the manuscript.  This publication makes use of data products
from the Two Micron All Sky Survey, which is a joint project of the
University of Massachusetts and the Infrared Processing and Analysis
Center/California Institute of Technology, funded by the National
Aeronautics and Space Administration and the National Science Foundation.
This publication also makes use of the Digital Sky Survey (POSS plates),
based on photographic data of the National Geographic Society -- Palomar
Geographic Society to the California Institute of Technology. The plates
were processed into the present compressed digital form with their
permission. The Digitized Sky Survey was produced at the Space Telescope
Science Institute under US Government grant NAG W-2166.  SJK acknowledges
support from a NASA GSRP fellowship.

\newpage

\appendix

\section{Tully-Fisher Fitting Technique}
\label{sc:fitting}

TF samples are generally incomplete at low luminosities and low
surface brightnesses, resulting in asymmetric scatter about the
fundamental relation.  In a fit to magnitude as a function of $\log v$
(the ``forward'' TFR), this asymmetric scatter leads to artificially
shallow slopes.  Extended luminosity coverage, as in the NFGS, helps
to alleviate the problem, but asymmetric faint-end scatter can still
bias results.

\citet{willick:statistical} describes one way to address TF slope
bias, using an iterative analysis.  His technique assumes that the
``missing'' galaxies have the same intrinsic TF slope and scatter as
the observed galaxies.  However, if either the scatter or the slope
increases at low luminosity, the bias-corrected slope is still too
shallow.

Alternatively, one can fit the ``inverse'' relation (minimizing the
scatter in velocity width as a function of magnitude) without any bias
correction, assuming that velocity width biases are negligible
\citep[cf.][]{tully.pierce:distances,schechter:mass-to-light}.  This
approach seems appropriate for our sample, because: (1) optical RC
data (unlike radio data) are not subject to a velocity width detection
bias, and (2) the NFGS makes no explicit diameter cut that might
introduce an indirect velocity width bias through a correlation of
rotation speed with surface brightness.  However, intrinsic surface
brightness and color bias in the parent survey may still play a role.

Tables~1 and 2 present our basic TF calibrations using three different
techniques: error-weighted inverse fits, bivariate fits with
regression in both magnitude and velocity width errors, and unweighted
inverse fits, our preferred technique.  Error-weighted inverse fits
avoid the usual TF slope bias, but they favor galaxies with higher
luminosity and larger velocity width, which have smaller errors.
Bivariate fits use all the available information but at the expense of
introducing some slope bias \citep[c.f.\ ][]{sakai.mould.ea:hubble}.
Note that the systematic differences between these fitting techniques
often exceed the formal errors (Tables~1 and 2).  

We compute unweighted inverse fits in two steps.  First we find the
68\% confidence interval for the slope, where for each test value of
the slope we allow the intercept to assume the value that minimizes
deviations for that slope.  Once we have the best-fit slope, we fit
for the intercept with the slope fixed.  Although the results are the
same as those of a one-step two-parameter fit, the errors are
different.  This type of fit is appropriate for comparing the slopes
of different samples irrespective of zero point, or conversely for
comparing their zero points at fixed slope.

All fits are subject to one round of 3$\sigma$ rejection.  Such
rejection reduces measured scatter in a Gaussian distribution by
$\sim$2\%, which we correct for in $\sigma_{3\sigma\, \rm rej}$.  Using
the same slope and zero point we also compute the full scatter
$\sigma_{\rm all\, pts}$, the biweight scatter
\citep{beers.flynn.ea:measures}, and the predicted scatter from
observational errors $\sigma_{\rm pred}$.  The biweight statistic
provides the most robust measure of observed scatter, generally
agreeing closely with $\sigma_{3\sigma\, \rm rej}$.

\section{Velocity Width Definitions \& Optical-To-Radio Conversions}
\label{sc:vdefns}

Below we present three different optical velocity width definitions,
$V_{max}$, $V_{fit}$, and $V_{pmm}$, and place them on an equivalent
W$_{50}$ scale.  Our optical-to-radio conversions are expected to differ
from those of \citet{courteau:optical}, not just because of differences in
sample selection, but also because Courteau's conversions are calibrated
using turbulence-corrected H\,{\small I} linewidths \citep[taken
from][]{giovanelli.haynes.ea:i}.  In our case, the conversions are
calibrated using uncorrected H\,{\small I} linewidths.

\subsection{$V_{max}$}
\label{sc:vmaxdefn}

\begin{figure}[tb]
\epsscale{0.65}
\plotone{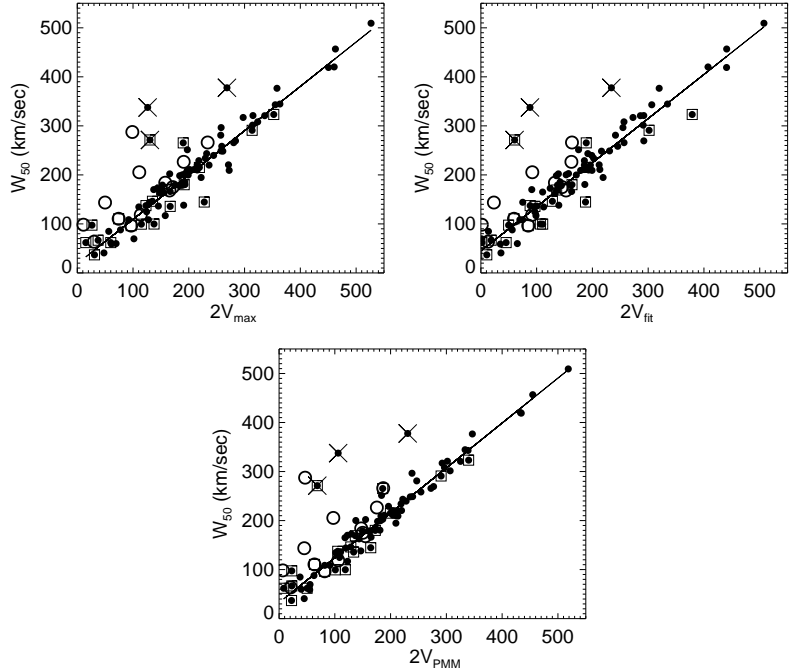}
\caption{W$_{50}$ versus $V_{max}$, $V_{fit}$, and $V_{pmm}$ for the 96
galaxies with both optical RC and H\,{\small I} W$_{50}$ velocity widths.
Boxes indicate points for which the optical rotation curve does not extend
to 1.3$r_{e}$; remarkably, these points show no strong deviation from the
general correlation.  Lines indicate least squares fits to the data with
automatically rejected points ($>$3.5$\sigma$) marked by an X.  All three
X's represent galaxies currently interacting with another galaxy: NGC~4795,
NGC~7752, and NGC~2799.  The open-circle points mark galaxies with
potentially large position angle misalignment in the optical RC data
(\S\ref{sc:orcobs}); these are also excluded from the fit.  The zero point
offset indicates the nonzero contribution of turbulence to the H\,{\small
I} profiles even at zero rotation velocity.}
\label{fg:w50vsopt}
\end{figure}

We define $V_{max}$ as the single largest velocity in the rotation curve,
relative to the origin defined by minimizing the RC asymmetry
(Appendix~\ref{sc:asymmeast}).  Because this number does not make use of
the full information in the rotation curve, we expect it to be somewhat
unreliable.  Nonetheless, it correlates surprisingly well with the radio
linewidth W$_{50}$ (Figure~\ref{fg:w50vsopt}), even when the optical RC
does not extend as far as 1.3$r_{e}$ \citet[the peak velocity position for
an exponential disk with effective radius $r_e$,][]{freeman:on}.  An
iterative least-squares fit with automatic 3.5$\sigma$ outlier rejection
yields
\begin{equation}
\rm W_{50} = 19\left(\pm6\right) + 
	0.90\left(\pm0.03\right)\left(2V_{max}\right)
\end{equation}
with scatter 25 $\rm km \, s^{-1}$.

\subsection{$V_{fit}$}
\label{sc:vfitdefn}

Following \citet{courteau:optical}, we define $V_{fit}$ by first
fitting the empirical function 
\begin{equation}
v=v_{0}+v_{c}\frac{(1+x)^{\beta}}{(1+x^{\gamma})^{1/\gamma}}
\end{equation}
to the observed RC, with the origin unconstrained, and then
interpolating to find the velocity at a specific radius.  Here
$x=r_{t}/(r-r_{0})$, $(r_{0},v_{0})$ defines the origin, $v_{c}$ gives
the velocity scale, $r_{t}$ is related to the turnover radius, and
$\gamma$ and $\beta$ are free parameters governing the shape of the
rotation curve.  We modify Courteau's technique slightly, in that we
interpolate the velocity $V_{fit}$ at 1.3$r_{e}$ rather than at
2.2 disk scale lengths $r_d$, because 1.3$r_{e}$ remains well defined
even for non-disk galaxies and does not require bulge+disk
decomposition.  Theoretically the two radii are equivalent, in that
both represent the peak velocity position for a pure exponential disk
\citep[][]{freeman:on}.  For disk-dominated galaxies in the NFGS,
1.3$r_{e}$ and 2.2$r_d$ also match well observationally. (Disk scale
lengths are courtesy of R. Jansen, private communication; both disk
scale lengths and $r_{e}$'s are from the B band images of
\citet{jansen.franx.ea:surface}.)

As shown in Figure~\ref{fg:rcgallery}, our modified-Courteau technique
performs well for bright spiral galaxies comparable to those in Courteau's
sample, but less regular RCs are poorly modeled.  Figure~\ref{fg:w50vsopt}
displays the correlation of $V_{fit}$ with the radio linewidth W$_{50}$,
demonstrating reasonable agreement, although some points are simply missing
because the fit would not converge.  An iterative least-squares fit with
automatic 3.5$\sigma$ outlier rejection yields
\begin{equation}
\rm W_{50} = 45\left(\pm5\right) + 
	0.90\left(\pm0.03\right)\left(2V_{fit}\right).
\end{equation}
We define the equivalent W$_{50}$ linewidth W$_{V_{fit}}$ by this
equation and assign it an error equal to the scatter in the fit, 25
$\rm km \, s^{-1}$.

\subsection{$V_{pmm}$}
\label{sc:vpmmdefn}

Following \citet{raychaudhury..ea:tests}, we define $V_{pmm}$ as
half the difference between the statistical ``probable maximum'' and
``probable minimum'' velocities implied by the observed rotation
curve.  The probable maximum velocity $v_{pmax}$ is defined to have a
10 percent chance of exceeding all velocities in the rotation curve:
\begin{equation}
\prod_{i} P \left( v_{pmax} > v_{i} \right) = 0.1.
\end{equation}
Modeling each velocity $v_{i}$ as a Gaussian distribution about the
measured value, with $\sigma$ equal to the measurement error, we obtain
\begin{equation}
\prod_{i} \left( \frac {1} {2} +\frac {1} {2} {\rm erf} {\frac
{v_{pmax} -v_{i}} {\sqrt 2\sigma_{i}}} \right) = 0.1,
\end{equation}
which may be solved numerically for $v_{pmax}$.  The probable minimum
velocity $v_{pmin}$ is defined by analogy, and $2V_{pmm} = v_{pmax}
-v_{pmin}$.

\citet{courteau:optical} suggests that $V_{pmm}$ will be very sensitive to
outlying points in the rotation curve, while \citet{raychaudhury..ea:tests}
claim the opposite.  The difference seems to hinge on the size of the error
bars associated with the outliers --- if a high S/N cosmic ray hit
masquerades as a very high-velocity data point, it may indeed throw off
$V_{pmm}$, but outlying points with large error bars will have relatively
little impact.  In practice, $V_{pmm}$ produces smaller scatter than
$V_{fit}$ in the optical-to-H\,{\small I} conversion.  More importantly, as
\citeauthor{raychaudhury..ea:tests} point out, $V_{pmm}$ has the virtue of
using all of the data in the rotation curve without imposing any particular
model on the data.  The diversity of the NFGS defies modeling through a
simple fitting function (see Figure~\ref{fg:rcgallery}).

Figure~\ref{fg:w50vsopt} shows the correlation of $V_{pmm}$ with
radio linewidth W$_{50}$.  An iterative least-squares fit with
automatic 3.5$\sigma$ outlier rejection yields
\begin{equation}
\rm W_{50} = 33\left(\pm5\right) + 
	0.92\left(\pm0.02\right)\left(2V_{pmm}\right).
\end{equation}
We define the equivalent W$_{50}$ linewidth W$_{V_{pmm}}$ by this
equation and assign it an error equal to the scatter in the fit, 20
$\rm km \, s^{-1}$.

\section{Rotation Curve Asymmetry Measurements}
\label{sc:asymmeast}

We compute optical rotation curve asymmetries using a technique
similar to the photometric technique of \citet{abraham.tanvir.ea:galaxy}.
Reflecting the rest-frame rotation curve about its origin, we find the
average absolute deviation between the two sides, $<\mid\!v-v^{\rm
reflected}\!\mid>$, and express the result as a percentage of the velocity
width $2V_{pmm}$ (Appendix~\ref{sc:vpmmdefn}).  Our asymmetry, velocity
width, and central redshift measurements are not completely independent of
one another, therefore in practice we perform these calculations
iteratively.

Asymmetry measurements depend sensitively on the choice of origin for
the spatial and velocity axes.  We therefore vary the position of the
origin to minimize the asymmetry, keeping the spatial coordinate
constrained within the one-sigma error bars of the continuum peak
position, while allowing the velocity coordinate to vary freely.  At
each step in the minimization process we resample the rotation curve
onto a uniform spatial grid about the trial origin.

To partially standardize our asymmetry index for rotation curves of
differing spatial extent, we define an inner asymmetry using only points
within 1.3$r_{e}$ (the peak velocity position for a pure exponential disk,
Appendix~\ref{sc:vfitdefn}).  When the rotation curve does not extend this
far, we compute the inner asymmetry from the data available. The rotation
curve origin is determined by minimizing the inner asymmetry.  All other
analysis makes use of the total asymmetry, which we compute relative to the
origin defined by minimizing the inner asymmetry.


\newpage

\begin{thebibliography}{86}
\expandafter\ifx\csname natexlab\endcsname\relax\def\natexlab#1{#1}\fi

\bibitem[{{Aaronson} {et~al.}(1982){Aaronson}, {Huchra}, {Mould}, {Schechter},
  \& {Tully}}]{aaronson.huchra.ea:velocity}
{Aaronson}, M., {Huchra}, J., {Mould}, J., {Schechter}, P.~L., \& {Tully},
  R.~B. 1982, \apj, 258, 64

\bibitem[{{Abraham} {et~al.}(1996){Abraham}, {Tanvir}, {Santiago}, {Ellis},
  {Glazebrook}, \& {van den Bergh}}]{abraham.tanvir.ea:galaxy}
{Abraham}, R.~G., {Tanvir}, N.~R., {Santiago}, B.~X., {Ellis}, R.~S.,
  {Glazebrook}, K., \& {van den Bergh}, S. 1996, \mnras, 279, L47

\bibitem[{{Barton} {et~al.}(2001){Barton}, {Geller}, {Bromley}, {van Zee}, \&
  {Kenyon}}]{barton.geller.ea:tully-fisher}
{Barton}, E.~J., {Geller}, M.~J., {Bromley}, B.~C., {van Zee}, L., \& {Kenyon},
  S.~J. 2001, \aj, 121, 625

\bibitem[{{Barton} \& {van Zee}(2001)}]{barton.:possible}
{Barton}, E.~J. \& {van Zee}, L. 2001, \apjl, 550, L35

\bibitem[{{Beers} {et~al.}(1990){Beers}, {Flynn}, \&
  {Gebhardt}}]{beers.flynn.ea:measures}
{Beers}, T.~C., {Flynn}, K., \& {Gebhardt}, K. 1990, \aj, 100, 32

\bibitem[{{Bell} \& {de Jong}(2001)}]{bell.:stellar}
{Bell}, E.~F. \& {de Jong}, R.~S. 2001, \apj, 550, 212

\bibitem[{{Bershady} \& {Andersen}(2000)}]{bershady.andersen:evolution}
{Bershady}, M.~A. \& {Andersen}, D.~R. 2000, in ASP Conf. Ser. 197: Dynamics of
  Galaxies: from the Early Universe to the Present, 175

\bibitem[{{Bershady} {et~al.}(1999){Bershady}, {Haynes}, {Giovanelli}, \&
  {Andersen}}]{bershady.haynes.ea:rotation}
{Bershady}, M.~A., {Haynes}, M.~P., {Giovanelli}, R., \& {Andersen}, D.~R.
  1999, in ASP Conf. Ser. 182: Galaxy Dynamics - A Rutgers Symposium, 499

\bibitem[{{Bottinelli} {et~al.}(1990){Bottinelli}, {Gouguenheim}, {Fouque}, \&
  {Paturel}}]{bottinelli.gouguenheim.ea:extragalactic}
{Bottinelli}, L., {Gouguenheim}, L., {Fouque}, P., \& {Paturel}, G. 1990,
  \aaps, 82, 391

\bibitem[{{Buchalter} {et~al.}(2001){Buchalter}, {Jimenez}, \&
  {Kamionkowski}}]{buchalter.jimenez.ea:galactosynthesis}
{Buchalter}, A., {Jimenez}, R., \& {Kamionkowski}, M. 2001, \mnras, 322, 43

\bibitem[{{Bureau} {et~al.}(1996){Bureau}, {Mould}, \&
  {Staveley-Smith}}]{bureau.mould.ea:new}
{Bureau}, M., {Mould}, J.~R., \& {Staveley-Smith}, L. 1996, \apj, 463, 60

\bibitem[{{Courteau}(1996)}]{courteau:deep}
{Courteau}, S. 1996, \apjs, 103, 363

\bibitem[{{Courteau}(1997)}]{courteau:optical}
---. 1997, \aj, 114, 2402

\bibitem[{{Courteau} \& {Rix}(1999)}]{courteau.rix:maximal}
{Courteau}, S.~. \& {Rix}, H. 1999, \apj, 513, 561

\bibitem[{{Eisenstein} \& {Loeb}(1996)}]{eisenstein.loeb:can}
{Eisenstein}, D.~J. \& {Loeb}, A. 1996, \apj, 459, 432

\bibitem[{{Elizondo} {et~al.}(1999){Elizondo}, {Yepes}, {Kates}, {M{\"u}ller},
  \& {Klypin}}]{elizondo.yepes.ea:self-regulating}
{Elizondo}, D., {Yepes}, G., {Kates}, R., {M{\"u}ller}, V., \& {Klypin}, A.
  1999, \apj, 515, 525

\bibitem[{{Fabricant} {et~al.}(1998){Fabricant}, {Cheimets}, {Caldwell}, \&
  {Geary}}]{fabricant.cheimets.ea:fast}
{Fabricant}, D., {Cheimets}, P., {Caldwell}, N., \& {Geary}, J. 1998, \pasp,
  110, 79

\bibitem[{{Forbes} {et~al.}(1996){Forbes}, {Phillips}, {Koo}, \&
  {Illingworth}}]{forbes.phillips.ea:keck}
{Forbes}, D.~A., {Phillips}, A.~C., {Koo}, D.~C., \& {Illingworth}, G.~D. 1996,
  \apj, 462, 89

\bibitem[{{Franx} \& {de Zeeuw}(1992)}]{franx.:elongated}
{Franx}, M. \& {de Zeeuw}, T. 1992, \apjl, 392, L47

\bibitem[{{Freeman}(1970)}]{freeman:on}
{Freeman}, K.~C. 1970, \apj, 160, 811

\bibitem[{{Frei} \& {Gunn}(1994)}]{frei.gunn:generating}
{Frei}, Z. \& {Gunn}, J.~E. 1994, \aj, 108, 1476

\bibitem[{{Garnier} {et~al.}(1996){Garnier}, {Paturel}, {Petit}, {Marthinet},
  \& {Rousseau}}]{garnier.paturel.ea:image}
{Garnier}, R., {Paturel}, G., {Petit}, C., {Marthinet}, M.~C., \& {Rousseau},
  J. 1996, \aaps, 117, 467

\bibitem[{{Giovanelli} {et~al.}(1997{\natexlab{a}}){Giovanelli}, {Haynes},
  {Herter}, {Vogt}, {da Costa}, {Freudling}, {Salzer}, \&
  {Wegner}}]{giovanelli.haynes.ea:i*1}
{Giovanelli}, R., {Haynes}, M.~P., {Herter}, T., {Vogt}, N.~P., {da Costa},
  L.~N., {Freudling}, W., {Salzer}, J.~J., \& {Wegner}, G. 1997{\natexlab{a}},
  \aj, 113, 53

\bibitem[{{Giovanelli} {et~al.}(1997{\natexlab{b}}){Giovanelli}, {Haynes},
  {Herter}, {Vogt}, {Wegner}, {Salzer}, {da Costa}, \&
  {Freudling}}]{giovanelli.haynes.ea:i}
{Giovanelli}, R., {Haynes}, M.~P., {Herter}, T., {Vogt}, N.~P., {Wegner}, G.,
  {Salzer}, J.~J., {da Costa}, L.~N., \& {Freudling}, W. 1997{\natexlab{b}},
  \aj, 113, 22

\bibitem[{{Giraud}(1986)}]{giraud:two-color}
{Giraud}, E. 1986, \aap, 164, 17

\bibitem[{{Gordon} {et~al.}(1997){Gordon}, {Calzetti}, \&
  {Witt}}]{gordon.calzetti.ea:dust}
{Gordon}, K.~D., {Calzetti}, D., \& {Witt}, A.~N. 1997, \apj, 487, 625

\bibitem[{{Han} \& {Mould}(1990)}]{han.mould:velocity}
{Han}, M. \& {Mould}, J. 1990, \apj, 360, 448

\bibitem[{{Haynes} {et~al.}(1999){Haynes}, {Giovanelli}, {Chamaraux}, {da
  Costa}, {Freudling}, {Salzer}, \& {Wegner}}]{haynes.giovanelli.ea:i-band}
{Haynes}, M.~P., {Giovanelli}, R., {Chamaraux}, P., {da Costa}, L.~N.,
  {Freudling}, W., {Salzer}, J.~J., \& {Wegner}, G. 1999, \aj, 117, 2039

\bibitem[{{Haynes} {et~al.}(2000){Haynes}, {Jore}, {Barrett}, {Broeils}, \&
  {Murray}}]{haynes.jore.ea:kinematic}
{Haynes}, M.~P., {Jore}, K.~P., {Barrett}, E.~A., {Broeils}, A.~H., \&
  {Murray}, B.~M. 2000, \aj, 120, 703

\bibitem[{{Huchra} {et~al.}(1983){Huchra}, {Davis}, {Latham}, \&
  {Tonry}}]{huchra.davis.ea:survey}
{Huchra}, J., {Davis}, M., {Latham}, D., \& {Tonry}, J. 1983, \apjs, 52, 89

\bibitem[{{Isobe} {et~al.}(1990){Isobe}, {Feigelson}, {Akritas}, \&
  {Babu}}]{isobe.feigelson.ea:linear}
{Isobe}, T., {Feigelson}, E.~D., {Akritas}, M.~G., \& {Babu}, G.~J. 1990, \apj,
  364, 104

\bibitem[{{Jacoby} {et~al.}(1992){Jacoby}, {Branch}, {Clardullo}, {Davies},
  {Harris}, {Pierce}, {Pritchet}, {Tonry}, \&
  {Welch}}]{jacoby.branch.ea:critical}
{Jacoby}, G.~H., {Branch}, D., {Clardullo}, R., {Davies}, R.~L., {Harris},
  W.~E., {Pierce}, M.~J., {Pritchet}, C.~J., {Tonry}, J.~L., \& {Welch}, D.~L.
  1992, \pasp, 104, 599

\bibitem[{{Jansen}(2000)}]{jansen:nearby}
{Jansen}, R.~A. 2000, PhD thesis, Univ.\ Groningen, The Netherlands

\bibitem[{{Jansen} {et~al.}(2000{\natexlab{a}}){Jansen}, {Fabricant}, {Franx},
  \& {Caldwell}}]{jansen.fabricant.ea:spectrophotometry}
{Jansen}, R.~A., {Fabricant}, D., {Franx}, M., \& {Caldwell}, N.
  2000{\natexlab{a}}, \apjs, 126, 331

\bibitem[{{Jansen} {et~al.}(2000{\natexlab{b}}){Jansen}, {Franx}, {Fabricant},
  \& {Caldwell}}]{jansen.franx.ea:surface}
{Jansen}, R.~A., {Franx}, M., {Fabricant}, D., \& {Caldwell}, N.
  2000{\natexlab{b}}, \apjs, 126, 271

\bibitem[{{Jorgensen}(1994)}]{jorgensen:secondary}
{Jorgensen}, I. 1994, \pasp, 106, 967

\bibitem[{{Kannappan}(2001)}]{kannappan:kinematic}
{Kannappan}, S.~J. 2001, PhD thesis, Harvard University

\bibitem[{{Kannappan} \& {Fabricant}(2001)}]{kannappan.fabricant:broad}
{Kannappan}, S.~J. \& {Fabricant}, D.~G. 2001, \aj, 121, 140

\bibitem[{{Kennicutt}(1998)}]{kennicutt:star}
{Kennicutt}, R.~C., J. 1998, \araa, 36, 189

\bibitem[{{Kennicutt} {et~al.}(1994){Kennicutt}, {Tamblyn}, \&
  {Congdon}}]{kennicutt.tamblyn.ea:past}
{Kennicutt}, R.~C., J., {Tamblyn}, P., \& {Congdon}, C.~E. 1994, \apj, 435, 22

\bibitem[{{Kodaira} \& {Watanabe}(1988)}]{kodaira.watanabe:statistical}
{Kodaira}, K. \& {Watanabe}, M. 1988, \aj, 96, 1593

\bibitem[{{Kraan-Korteweg} {et~al.}(1984){Kraan-Korteweg}, {Sandage}, \&
  {Tammann}}]{kraan-korteweg.sandage.ea:effect}
{Kraan-Korteweg}, R.~C., {Sandage}, A., \& {Tammann}, G.~A. 1984, \apj, 283, 24

\bibitem[{{Marzke} {et~al.}(1994){Marzke}, {Huchra}, \&
  {Geller}}]{marzke.huchra.ea:luminosity}
{Marzke}, R.~O., {Huchra}, J.~P., \& {Geller}, M.~J. 1994, \apj, 428, 43

\bibitem[{{Mathewson} {et~al.}(1992){Mathewson}, {Ford}, \&
  {Buchhorn}}]{mathewson.ford.ea:southern}
{Mathewson}, D.~S., {Ford}, V.~L., \& {Buchhorn}, M. 1992, \apjs, 81, 413

\bibitem[{{Matthews} {et~al.}(1998){Matthews}, {van Driel}, \&
  {Gallagher}}]{matthews..ea:exploration}
{Matthews}, L.~D., {van Driel}, W., \& {Gallagher}, J.~S. 1998, \aj, 116, 2196

\bibitem[{{McGaugh} {et~al.}(2000){McGaugh}, {Schombert}, {Bothun}, \& {de
  Blok}}]{mcgaugh.schombert.ea:baryonic}
{McGaugh}, S.~S., {Schombert}, J.~M., {Bothun}, G.~D., \& {de Blok}, W. J.~G.
  2000, \apjl, 533, L99

\bibitem[{{Mihos} \& {Hernquist}(1996)}]{mihos.hernquist:gasdynamics}
{Mihos}, J.~C. \& {Hernquist}, L. 1996, \apj, 464, 641

\bibitem[{{Neistein} {et~al.}(1999){Neistein}, {Maoz}, {Rix}, \&
  {Tonry}}]{neistein.maoz.ea:tully-fisher}
{Neistein}, E., {Maoz}, D., {Rix}, H., \& {Tonry}, J.~L. 1999, \aj, 117, 2666

\bibitem[{{Nilson}(1973)}]{nilson:uppsala}
{Nilson}, P. 1973, Uppsala general catalogue of galaxies (Uppsala: Astronomiska
  Observatorium, 1973)

\bibitem[{{O'Neil} {et~al.}(2000){O'Neil}, {Bothun}, \&
  {Schombert}}]{oneil.bothun.ea:red}
{O'Neil}, K., {Bothun}, G.~D., \& {Schombert}, J. 2000, \aj, 119, 136

\bibitem[{{Pierce} \& {Tully}(1988)}]{pierce.tully:distances}
{Pierce}, M.~J. \& {Tully}, R.~B. 1988, \apj, 330, 579

\bibitem[{{Pierce} \& {Tully}(1992)}]{pierce.tully:luminosity-line}
---. 1992, \apj, 387, 47

\bibitem[{{Pierini}(1999)}]{pierini:internal}
{Pierini}, D. 1999, \aap, 352, 49

\bibitem[{{Pierini} \& {Tuffs}(1999)}]{pierini.tuffs:linear}
{Pierini}, D. \& {Tuffs}, R.~J. 1999, \aap, 343, 751

\bibitem[{{Pustilnik} {et~al.}(2001){Pustilnik}, {Kniazev}, {Lipovetsky}, \&
  {Ugryumov}}]{pustilnik.kniazev.ea:environment}
{Pustilnik}, S.~A., {Kniazev}, A.~Y., {Lipovetsky}, V.~A., \& {Ugryumov}, A.~V.
  2001, \aap, 373, 24

\bibitem[{{Raychaudhury} {et~al.}(1997){Raychaudhury}, {von Braun},
  {Bernstein}, \& {Guhathakurta}}]{raychaudhury..ea:tests}
{Raychaudhury}, S., {von Braun}, K., {Bernstein}, G.~M., \& {Guhathakurta}, P.
  1997, \aj, 113, 2046

\bibitem[{{Rhee}(1996)}]{rhee:physical}
{Rhee}, M.-H. 1996, PhD thesis, Univ.\ Groningen, The Netherlands

\bibitem[{{Rix} {et~al.}(1997){Rix}, {Guhathakurta}, {Colless}, \&
  {Ing}}]{rix.guhathakurta.ea:internal}
{Rix}, H., {Guhathakurta}, P., {Colless}, M., \& {Ing}, K. 1997, \mnras, 285,
  779

\bibitem[{{Roberts}(1978)}]{roberts:twenty-one}
{Roberts}, M.~S. 1978, \aj, 83, 1026

\bibitem[{{Rubin} {et~al.}(1985){Rubin}, {Burstein}, {Ford}, \&
  {Thonnard}}]{rubin.burstein.ea:rotation}
{Rubin}, V.~C., {Burstein}, D., {Ford}, W.~K., \& {Thonnard}, N. 1985, \apj,
  289, 81

\bibitem[{{Sakai} {et~al.}(2000){Sakai}, {Mould}, {Hughes}, {Huchra}, {Macri},
  {Kennicutt}, {Gibson}, {Ferrarese}, {Freedman}, {Han}, {Ford}, {Graham},
  {Illingworth}, {Kelson}, {Madore}, {Sebo}, {Silbermann}, \&
  {Stetson}}]{sakai.mould.ea:hubble}
{Sakai}, S., {Mould}, J.~R., {Hughes}, S. M.~G., {Huchra}, J.~P., {Macri},
  L.~M., {Kennicutt}, R.~C., {Gibson}, B.~K., {Ferrarese}, L., {Freedman},
  W.~L., {Han}, M., {Ford}, H.~C., {Graham}, J.~A., {Illingworth}, G.~D.,
  {Kelson}, D.~D., {Madore}, B.~F., {Sebo}, K., {Silbermann}, N.~A., \&
  {Stetson}, P.~B. 2000, \apj, 529, 698

\bibitem[{{Schechter}(1980)}]{schechter:mass-to-light}
{Schechter}, P.~L. 1980, \aj, 85, 801

\bibitem[{{Simard} \& {Pritchet}(1998)}]{simard.pritchet:internal}
{Simard}, L. \& {Pritchet}, C.~J. 1998, \apj, 505, 96

\bibitem[{{Stil} \& {Israel}(1998)}]{stil.israel:faint}
{Stil}, J.~M. \& {Israel}, F.~P. 1998, in astro-ph/9810151

\bibitem[{{Strauss} \& {Willick}(1995)}]{strauss.willick:density}
{Strauss}, M.~A. \& {Willick}, J.~A. 1995, \physrep, 261, 271

\bibitem[{{Taylor} {et~al.}(1996){Taylor}, {Thomas}, {Brinks}, \&
  {Skillman}}]{taylor.thomas.ea:survey}
{Taylor}, C.~L., {Thomas}, D.~L., {Brinks}, E., \& {Skillman}, E.~D. 1996,
  \apjs, 107, 143

\bibitem[{{Theureau} {et~al.}(1998){Theureau}, {Bottinelli}, {Coudreau-Durand},
  {Gouguenheim}, {Hallet}, {Loulergue}, {Paturel}, \&
  {Teerikorpi}}]{theureau.bottinelli.ea:kinematics}
{Theureau}, G., {Bottinelli}, L., {Coudreau-Durand}, N., {Gouguenheim}, L.,
  {Hallet}, N., {Loulergue}, M., {Paturel}, G., \& {Teerikorpi}, P. 1998,
  \aaps, 130, 333

\bibitem[{{Trentham} {et~al.}(2001){Trentham}, {Tully}, \&
  {Verheijen}}]{trentham.tully.ea:ursa}
{Trentham}, N., {Tully}, R.~B., \& {Verheijen}, M.~A.~W. 2001, \mnras, 325, 385

\bibitem[{{Tully}(1988)}]{tully:nearby}
{Tully}, R.~B. 1988, Nearby galaxies catalog (Cambridge and New York, Cambridge
  University Press)

\bibitem[{{Tully} \& {Fisher}(1977)}]{tully.fisher:new}
{Tully}, R.~B. \& {Fisher}, J.~R. 1977, \aap, 54, 661

\bibitem[{{Tully} \& {Fouque}(1985)}]{tully.fouque:extragalactic}
{Tully}, R.~B. \& {Fouque}, P. 1985, \apjs, 58, 67

\bibitem[{{Tully} \& {Pierce}(2000)}]{tully.pierce:distances}
{Tully}, R.~B. \& {Pierce}, M.~J. 2000, \apj, 533, 744

\bibitem[{{Tully} {et~al.}(1998){Tully}, {Pierce}, {Huang}, {Saunders},
  {Verheijen}, \& {Witchalls}}]{tully.pierce.ea:global}
{Tully}, R.~B., {Pierce}, M.~J., {Huang}, J., {Saunders}, W., {Verheijen}, M.
  A.~W., \& {Witchalls}, P.~L. 1998, \aj, 115, 2264

\bibitem[{{Tully} {et~al.}(1992){Tully}, {Shaya}, \&
  {Pierce}}]{tully.shaya.ea:nearby}
{Tully}, R.~B., {Shaya}, E.~J., \& {Pierce}, M.~J. 1992, \apjs, 80, 479

\bibitem[{{Tully} {et~al.}(1996){Tully}, {Verheijen}, {Pierce}, {Huang}, \&
  {Wainscoat}}]{tully.verheijen.ea:ursa}
{Tully}, R.~B., {Verheijen}, M. A.~W., {Pierce}, M.~J., {Huang}, J., \&
  {Wainscoat}, R.~J. 1996, \aj, 112, 2471

\bibitem[{{Tutui} \& {Sofue}(1997)}]{tutui.sofue:effects}
{Tutui}, Y. \& {Sofue}, Y. 1997, \aap, 326, 915

\bibitem[{{Verheijen}(1997)}]{verheijen:ursa}
{Verheijen}, M. A.~W. 1997, PhD thesis, Univ.\ Groningen, The Netherlands

\bibitem[{{Verheijen}(2001)}]{verheijen:ursa*1}
{Verheijen}, M.~A.~W. 2001, \apj, accepted

\bibitem[{{Verheijen} \& {Sancisi}(2001)}]{verheijen.sancisi:ursa}
{Verheijen}, M.~A.~W. \& {Sancisi}, R. 2001, \aap, 370, 765

\bibitem[{{Vogt} {et~al.}(1997){Vogt}, {Phillips}, {Faber}, {Gallego},
  {Gronwall}, {Guzman}, {Illingworth}, {Koo}, \&
  {Lowenthal}}]{vogt.phillips.ea:optical}
{Vogt}, N.~P., {Phillips}, A.~C., {Faber}, S.~M., {Gallego}, J., {Gronwall},
  C., {Guzman}, R., {Illingworth}, G.~D., {Koo}, D.~C., \& {Lowenthal}, J.~D.
  1997, \apjl, 479, L121

\bibitem[{{Watanabe} {et~al.}(2001){Watanabe}, {Yasuda}, {Itoh}, {Ichikawa}, \&
  {Yanagisawa}}]{watanabe.yasuda.ea:surface}
{Watanabe}, M., {Yasuda}, N., {Itoh}, N., {Ichikawa}, T., \& {Yanagisawa}, K.
  2001, \apj, 555, 215

\bibitem[{{Willick}(1994)}]{willick:statistical}
{Willick}, J.~A. 1994, \apjs, 92, 1

\bibitem[{{Willick} {et~al.}(1996){Willick}, {Courteau}, {Faber}, {Burstein},
  {Dekel}, \& {Kolatt}}]{willick.courteau.ea:homogeneous}
{Willick}, J.~A., {Courteau}, S., {Faber}, S.~M., {Burstein}, D., {Dekel}, A.,
  \& {Kolatt}, T. 1996, \apj, 457, 460

\bibitem[{{Yasuda} {et~al.}(1997){Yasuda}, {Fukugita}, \&
  {Okamura}}]{yasuda.fukugita.ea:study}
{Yasuda}, N., {Fukugita}, M., \& {Okamura}, S. 1997, \apjs, 108, 417

\bibitem[{{Zaritsky}(1995)}]{zaritsky:evidence}
{Zaritsky}, D. 1995, \apjl, 448, L17

\bibitem[{{Zaritsky} \& {Rix}(1997)}]{zaritsky.rix:lopsided}
{Zaritsky}, D. \& {Rix}, H. 1997, \apj, 477, 118

\end{thebibliography}
\end{document}